\documentclass[twocolumn,showpacs,prl,english,showkeys,10pt,aps,nofootinbib]{revtex4-1}

\usepackage[utf8]{inputenc}
\usepackage{graphicx,amsmath,amssymb}
\usepackage[english]{babel}
\usepackage{braket}

\newcommand{\bx}{{\boldsymbol x}}
\newcommand{\by}{{\boldsymbol y}}
\newcommand{\br}{{\boldsymbol r}}

\newcommand{\SG}{\mathrm{SG}}

\newcommand{\RF}{\mathrm{RF}}
\newcommand{\F}{\mathrm{F}}
\newcommand{\amp}{\mathrm{amp}}

\newcommand{\normauno}[1]{\left\|{#1}\right\|_1}

\newcommand{\norm}[1]{\left| {#1} \right|}

\newcommand{\M}{{\mathcal{M}}}
\newcommand{\IS}{\mathrm{IS}}
\newcommand{\hbe}{\hat{\mathrm{e}}}
\newcommand{\RAND}{\mathrm{RAND}}

\begin{document}

\title{Soft Modes, Localization and Two-Level Systems in Spin Glasses}
\author{M. Baity-Jesi} 
\affiliation{Departamento de F\'\i{}sica Te\'orica I, Universidad Complutense, 28040 Madrid, Spain,} 
\affiliation{Dipartimento di Fisica, La Sapienza Universit\`a di Roma, 00185 Roma,  Italy,}
\affiliation{Instituto de Biocomputaci\'on y F\'{\i}sica de Sistemas Complejos (BIFI), 50009 Zaragoza, Spain.}
\author{V. Mart\'{\i}n-Mayor}
\affiliation{Departamento de F\'\i{}sica Te\'orica I, Universidad Complutense, 28040 Madrid, Spain,} 
\affiliation{Instituto de Biocomputaci\'on y F\'{\i}sica de Sistemas Complejos (BIFI), 50009 Zaragoza, Spain.}
\author{G. Parisi}
\affiliation{Dipartimento di Fisica, NANOTEC-CNR, Soft and Living Matter Laboratory and INFN, La Sapienza Universit\`a di Roma, 00185 Roma,  Italy.}
\author{S. Perez-Gaviro}
\affiliation{Centro Universitario de la Defensa, Carretera de Huesca s/n, 50090 Zaragoza, Spain,}
\affiliation{Instituto de Biocomputaci\'on y F\'{\i}sica de Sistemas Complejos (BIFI), 50009 Zaragoza, Spain.}

\date{\today}

\begin{abstract}
In the three-dimensional Heisenberg spin glass in a random field we study the
properties of the inherent structures that are obtained by an instantaneous
cooling from infinite temperature.  For not too large field the density of
states $g(\omega)$ develops localized soft {\em plastic} modes and reaches
zero as $\omega^4$ (for large fiels a gap appears).  When we perturb the
system adding a force along the softest mode one reaches very similar minima
of the energy, separated by small barriers, that appear to be good candidates
for classical two-level systems.
\end{abstract}

 \maketitle

Supercooled liquids and amorphous solids exhibit an excess of
low-energy excitations, compared with their crystalline counterparts
\cite{phillips:81}, in which at low frequencies the density of states
(DOS) $g(\omega)$ has a Debye behavior $g(\omega)\propto\omega^{d-1}$
 in $d$ spatial dimensions.  This excess of low-frequency modes is
called \emph{Boson peak} \cite{buchenau:84,malinovsky:91} and it is
located at a small, but non zero frequency.

What happens at much lower frequency? Obviously we find phonons, but
what is there beyond phonons? Were it possible to disregard Goldstone
bosons, a scaling $g(\omega)\propto\omega^{\delta}$, with $\delta=3$
or 4 has been suggested for disordered systems
\cite{gurarie:03,gurevich:03}.  Still, this has not been demonstrated
nor observed. It has been stressed by \cite{hentschel:11,lin:15,xu:15}
that there are localized plastic modes, whose spectral density reaches
zero when $\omega$ goes to zero. These modes are subdominant in the
small frequency region: They are called ``{\sl plastic}'' because they
dominate the plastic response. These extra small frequency modes may
be related to the behaviour of hard spheres at jamming
\cite{wyart:05,wyart:12,ohern:03,franz:15,franz:15b}.

Replica theory offers an explanation for these extra modes. At low
enough temperatures, strongly disordered mean field models undergo
spontaneous full replica symmetry breaking (RSB).  Full RSB implies a
complex energy landscape with a hierarchical structure of states and a
large amount of degenerate minima separated by small free energy
barriers \cite{mezard:84,charbonneau:14}. As a consequence,
zero-temperature equilibrium configurations can be deformed at
essentially no energy cost through easy-deformation patterns, that we
name soft modes \cite{mezard:87,franz:15b}.  These modes are localized
in space, but non-exponentially. In fact, the zero temperature phase
transition from the replica symmetric phase to the RSB phase is
accompanied by a divergence of the localization length \cite{lupo:15}.

More often than not, finding low-lying energy minima of glassy systems
is a NP problem \cite{barahona:82b}.  Here we study the behavior of
inherent structures (IS), local minima of the energy obtained by
relaxing the system from high temperature (the thermal protocol should
not change drastically the DOS, at least if we remain in the replica
symmetric phase, see the appendix). In our study we need a model
with continuous degrees of freedom. The Heisenberg model, where the
spins are three-dimensional unitary vectors, is an epitome of the spin
glass \cite{anderson:70}.

The global rotational-symmetry of the Heisenberg spin glass has
far-reaching consequences. The corresponding symmetry in structural
glasses is translation symmetry (which has similar
implications). The Goldstone mechanism induces soft excitations in the
form of spin waves~\cite{bray:81}. Even in disordered systems,
spin-waves are efficiently labelled by their wavevector, especially at
low frequencies (see e.g. \cite{grigera:11}). As a
consequence, we have a Debye spectrum $g(\omega)\propto\omega^{d-1}$,
i.e. extended spin waves (for structural glasses the situation is slightly more complicated.
\footnote{
In the case of structural glasses, the authors of  
\cite{wyart:05b,lerner:13,degiuli:14b,degiuli:15} find $g(\omega)\sim \omega^2$ is valid for any spatial dimension. 
The origin of this different behaviour should be investigated carefully.
})
These symmetry-induced
modes mask the physics we aim to investigate.

Thus, we add a random magnetic field (RF) to wipe out the symmetries.
Indeed, mean field suggests that, if small enough, our RF does not destroy
the glass phase \cite{sharma:10}. In a RF, spin waves have a positive
frequency, even for vanishing wavevector (e.g. a ferromagnet in a
RF has no soft-modes). Therefore, the RF exposes the (possibly
localized) plastic modes that interest us. The resulting spectrum has
no reason to be Debye as it does not result from plain waves. Yet, a
crossover to the Debye regime should appear when the RF is small.  A
similar procedure of symmetry removal has been carried through in
glass-forming liquids, by pinning a certain fraction of particles
\cite{kob:12,cammarota:13,brito:13,gokhale:14,hima-nagamanasa:15}. The Heisenberg spin offers the
advantage of allowing us to simulate unprecedentedly large systems,
letting us observe scalings of several orders of magnitude.

Here, we study the ISs starting from initial random configurations 
and we do find that they are marginally stable states:
the distribution of eigenvalues of the Hessian matrix stretches down to
zero as a power law, and it is unrelated to symmetries in the system. Furthermore, we find that the soft modes are localized.
We also take in account the anharmonic effects due to the complexity of the
energy landscape. We find that the energy barriers along the softest mode are
extremely small and that they connect very similar states with an strong
relationship, that we propose as a operational definition of classical
two-level systems (TLS) \cite{grzonka:84}.

\paragraph{Model}
We study the three-dimensional Heisenberg spin glass in a RF.  The dynamic
variables are three-dimensional spins $\vec{s}_\bx$, placed at the vertices
$\bx$ of a cubic lattice of linear size $L$ with unitary spacings. We have
therefore $N=L^3$ spins, and $2N$ degrees of freedom (dof) due to the
normalization constraint $\vec{s}_\bx^{\,2}=1$.  The Hamiltonian is
\begin{equation}\label{eq:HRF}
 \mathcal{H}_\RF(\ket{\vec s}) = -\sum_{|\bx - \by|=1} J_{\bx\by} \vec{s}_{\bx}\cdot \vec{s}_{\by} - \sum_{\bx}^{N} \vec{h}_{\bx} \cdot \vec{s}_{\bx}\,,
\end{equation}
where the fields $\vec{h}_{\bx}$ are random vectors chosen uniformly from the
sphere of radius $H_\amp$, and $\ket{\vec s}$ indicates the full configuration
of spins $\vec s_\bx$.  The RF breaks all rotational and translational
symmetry, removing the Goldstone bosons.  The couplings $J_{i j}$ are fixed,
Gaussian distributed, with $\overline{J_{\bx\by}}=0$ and
$\overline{J_{\bx\by}^2}=J^2$, where $\overline{\left(\ldots\right)}$ is the
average over the disorder.

We simulated on systems of linear lattice size $L = 12, 24, 48, 96, 192$.  We
chose always $J=1$, and we compared it with $H_\amp = 0.01, 0.05, 0.1, 0.5,
1, 5, 10, 50$.  In the appendix we summarize the simulation
parameters. The case $H_\amp=0$ will be treated in a future work \cite{baityjesi:16},
because it requires a different type of analysis, since the spin waves do not
hybridize with the bulk of the spectrum.

\paragraph{Density of states}
We calculate the dynamical matrix as the Hessian matrix $\M$ of Hamiltonian
\eqref{eq:HRF}, calculated at the ISs.  In the appendix we report how the ISs were
obtained, we motivate the choice of the algorithm and the temperature of the
starting configuration $\ket{\vec s}$ for the energy minimizations, and we
show how the Hessian matrix $\M$ was calculated.

Once $\M$ is known, from each simulated $H_\amp$ we calculate the spectrum of
the eigenvalues $\rho(\lambda)$ or equivalently, in analogy with plane waves
\cite{huang:87}, the DOS $g(\omega)$, by defining $\lambda=\omega^2$.  We
measure the DOS both with
the method of the moments \cite{chihara:78,turchi:82,alonso:01}, and by 
explicitly computing with Arpack \cite{arpack} the lowest eigenvalues.
\footnote{The method of the moments returns the full density of
  states, but it is not precise at the tails. With Arpack we can
  calculate exactly the smallest eigenvalues, but only a small number
  of them. So, when we want to show the whole spectrum we need to use
  the method of the moments, while when we show the softest part of
  the spectrum we need Arpack.}

\begin{figure}[!t]
 \includegraphics[width=\columnwidth]{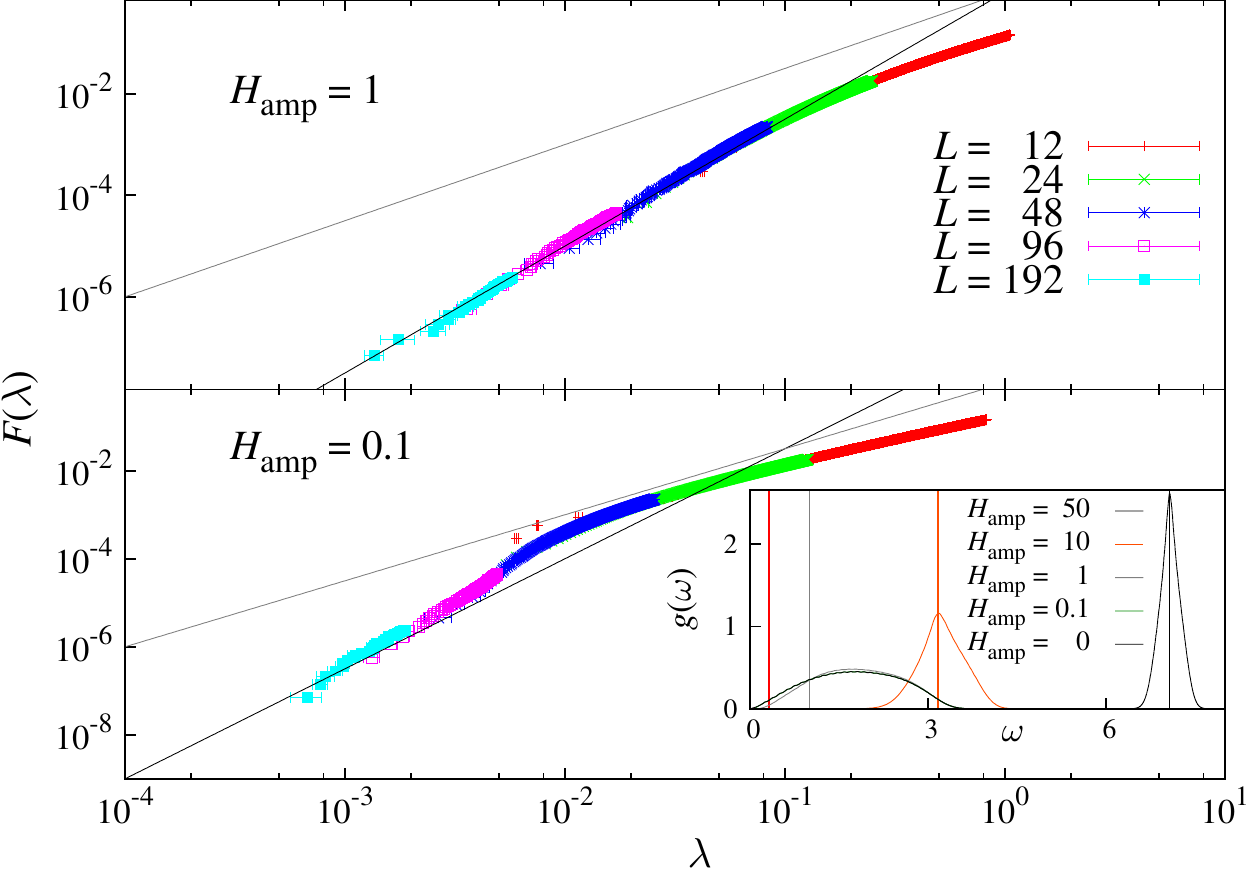}
 \caption{Cumulative $F(\lambda)$ of the spectrum of $\M$. In each plot we
   show black a reference curve representing the power law $\lambda^{2.5}$,
   and a grey line indicating the Debye behavior $\lambda^{1.5}$.
   \emph{Inset}: The DOS $g(\omega)$ calculated with the method of the
   moments. In the limit of a diagonal
   hamiltonian ($J=0$) the DOS would be a delta function centered on $\omega^2=H_\amp$.
   This value is represented with vertical lines.}
  \label{fig:cumulativas_inset}
\end{figure}

We find that, although for large fields there is a gap in the DOS when the
field is small enough the gap disappears and the DOS goes to zero developing
soft modes (Fig.~\ref{fig:cumulativas_inset}, inset).

We focus on the $\rho(\lambda)$ for small $\lambda$, or even better on its
cumulative function $F(\lambda) = \int_0^\lambda \rho(\lambda')d\lambda'$.  If
$F(\lambda)$ reaches zero as a power law, we can define three exponents
$\delta,\alpha$ and $\gamma$, that describe how the functions $g,\rho$ and $F$
go to zero for small $\lambda$ (or $\omega$):
\begin{equation}\label{eq:hsgrf-exponents}
  g(\omega)    \sim\omega^\delta\,,~~~
  \rho(\lambda)\sim\lambda^\alpha\,,~~~
  F(\lambda)   \sim\lambda^\gamma\,,
\end{equation}
where the exponents are related by $\delta = 2\alpha+1 = 2\gamma-1$.
In the absence of a field one expects a Debye-like behavior $\delta=d-1=2$, $\alpha=0.5$, $\gamma=1.5$ \cite{grigera:11}.

In Fig.~\ref{fig:cumulativas_inset} we show the function $F(\lambda)$ for fields $H_\amp=0.1, 1$. 
The plots are compared with the Debye behavior $\lambda^{1.5}$ and with the power law behavior $\lambda^{2.5}$,  because
our data suggests a universal behavior around $\gamma=2.5$ ($\delta=4$, $\alpha=1.5$) for all $H_\amp$ that does not
exhibit a gap.
\footnote{The value $\gamma=2.5$ is also hypothesized in \cite{gurarie:03}, through a fourth-order expansion of a single 
coordinate potential of the minimum of the energy.}
See the appendix for the data on other $H_\amp$.

When the field is small we remark a change of trend from
$\gamma\approx2.5$ to $\gamma<1.5$ at a value $\lambda^*$.  Very roughly
speaking, the crossover point goes as $\lambda^*\sim H_\amp^{-1}$,
maybe indicating the presence of a boson peak.

\begin{figure}[!t]
  \includegraphics[width=\columnwidth]{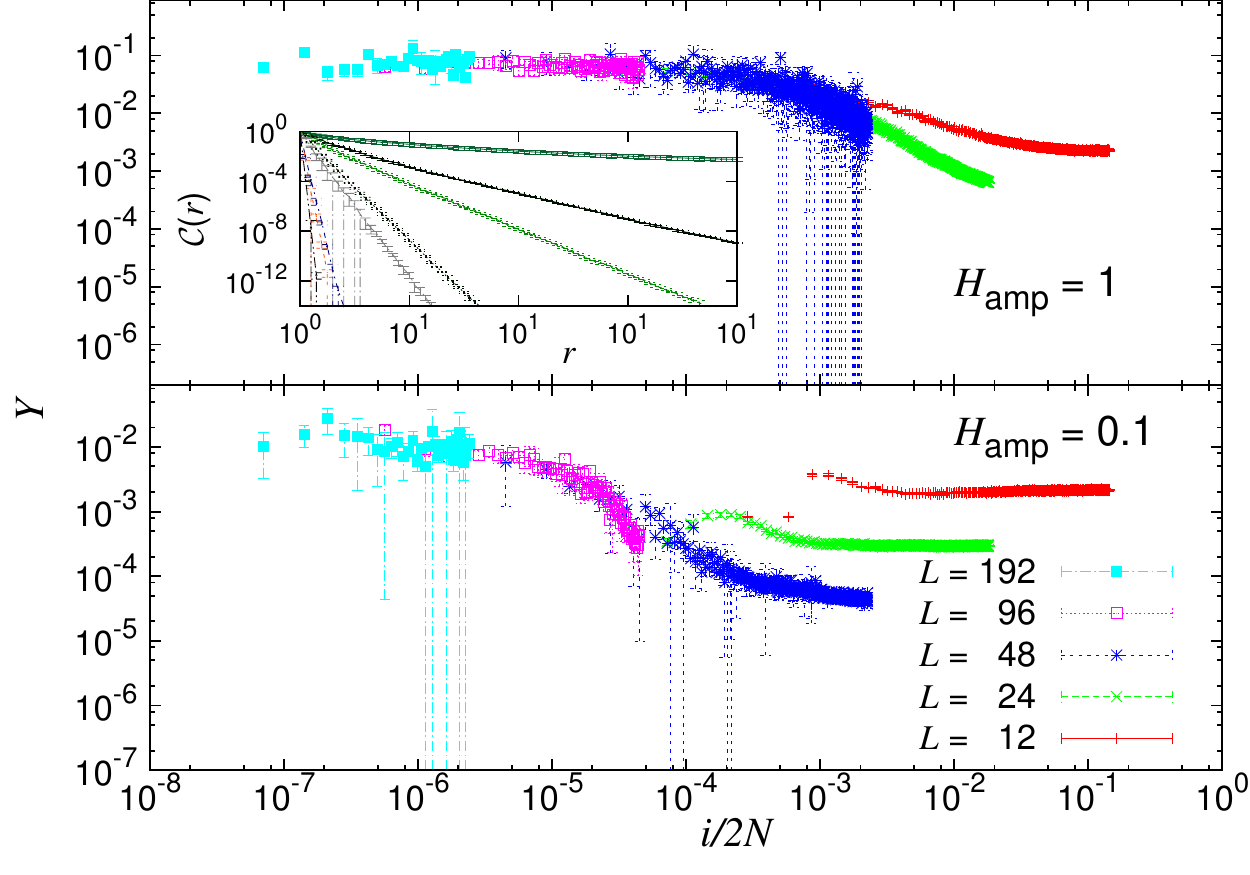}
 \caption{Inverse participation ratio as a function of the normalized
   rank $i/2N$ of the eigenvector ($i=1$ has the smallest eigenvalue,
   $i=2$ the second smallest, ...), for $H_\amp=1$ ({\bf top}) and
   $H_\amp=0.1$ ({\bf bottom}). {\bf Inset:} Correlation function
   $\mathcal{C}(r)$ extracted from the eigenvectors, for fields (from
   top to bottom) $H_\amp = 0.01, 0.05, 0.1, 0.5, 1, 5, 10, 50$ in
   $L=192$ lattices. See the appendices for a close-up. For the
   smallest $H_\amp$ our data do not display an exponential decay,
   which could be caused by a localization length larger than the
   system size.}
 \label{fig:local}
\end{figure}

\paragraph{Localization}
Similarly as it happens in other types of disordered systems
\cite{xu:10,degiuli:14,charbonneau:15}, the soft modes are localized,
meaning that the eigenvectors $\ket{\vec \pi_{n}}$ are dominated by
few components. A nice localization probe is the inverse participation
ratio $Y_n =
\frac{\sum_{\bx}(|{\vec\pi_{n,\bx}|^2)^2}}{(\sum_{\bx}|\vec\pi_{n,\bx}|^2)^2}$.
If the eigenvector $\ket{\pi_{n}}$ is fully localized in one site,
then $Y_n = 1$, whereas if it is fully delocalized, $Y_n = 1/N$. In
Fig.~\ref{fig:local} we show that the softer the eigenvectors the more
localized they are, and for infinitely large systems there is probably
a localization threshold that separates a small fixed percentage of
localized eigenvectors from the delocalized bulk ones.

The localization length increases as $H_\amp$ decreases, see
Fig.~\ref{fig:local}--inset. In fact, a RSB transition should cause a
localization transition at the critical $H_\amp$
\cite{lupo:15}. However, it is unclear whether or not an RSB
transition would leave a trace in infinite-temperature ISs.
\footnote{Inherent structures that were obtained by relaxing
an infinite-temperature configuration.}

\paragraph{Anharmonicity}
We go beyond the harmonic approximation, and take in account the relationship between different ISs.

We study the reaction of the system to a force along a direction
$\ket{\pi}$, normalized to one: $\sum_\bx \vec\pi_\bx^{\,2} = 1$.  We
examine  the softest mode, that is localized, and we compare it with
the behavior of the eigenvectors in bulk of the $\rho(\lambda)$, that
are delocalized. Therefore we choose $\ket{\pi}=\ket{\pi_0}$ (softest
mode) and $\ket{\pi}=\ket{\pi_\RAND}$, a vector whose components are
chosen at random.  The vector $\ket{\pi_\RAND}$ is not an eigenvector
of $\M$, but it is a random linear combination of all the eigenvectors of
the system. Since the bulk eigenvectors overwhelm the soft modes by
number, $\ket{\pi_\RAND}$ will be representative of the bulk behavior.

With the application of a forcing along $\ket{\vec\pi}$, Hamiltonian
\eqref{eq:HRF} is modified in
\begin{equation}\label{eq:ham-forcing}
 \mathcal{H}_\F(\ket{\vec s}) = -\sum_{\|\bx - \by\|=1} J_{\bx\by}
 \vec{s}_{\bx}\cdot \vec{s}_{\by} - \sum_{\bx}^{N}
 \left(\vec{h}_{\bx}+A_\F \vec{\pi}_{\bx}\right) \cdot
 \vec{s}_{\bx}\,,
\end{equation}
where $A_\F$ is the amplitude of the forcing. If $A_\F>0$ ($A_\F<0$),
spins tilt toward (against) $\ket{\vec\pi}$.  We can measure
quantitatively this response of the system to the forcing through
$\hat{m} = \sum_\bx \vec s_\bx\cdot\vec\pi_{\bx}$. We are interested
in forcings $A_\F$ both in the linear response regime, and just out of
it.

We stimulate the system with forcings of increasing amplitude, and
study when this kicks the system out of the original inherent
structure.  Ideally, the forcing amplitude $A_\F$ would grow
continuously. We simplify the analysis by choosing $A_\F = {\cal A}
i_h$, where ${\cal A}$ is a carefully tuned amplitude (see below)
while $i_h$ is an integer.  The unperturbed Hamiltonian corresponds to
$i_h=0$, while $i_h=\pm N_F$ is our maximum forcing (note that $\pm
i_h$ forcings are not equivalent due to anharmonicities).

This is how we check if new states were encountered upon increasing the forcing:
(\emph{i}) For each $i_h$, start from the IS
$\ket{\vec{s}^{\,(\IS)}}$ of the unperturbed Hamiltonian
$\mathcal{H}_\RF$.  (\emph{ii}) From
$\ket{\vec{s}^{\,(\IS)}}$ minimize the energy using
$\mathcal{H}_\F(i_h)$, and find a new IS for the perturbed system,
$\ket{\IS(i_h)}$.  (\emph{iii}) From $\ket{\IS(i_h)}$ minimize the
energy again, using $\mathcal{H}_\F(0)=\mathcal{H}_\RF$, and find the
IS $\ket{\IS^{*}}$ (with elements $\vec{s}^{\,(\IS)*}_\bx$).
(\emph{iv}) If $\ket{\IS^*}=\ket{\vec{s}^{(\IS)}}$, the second
minimization lead the system back to its original configuration, so
the forcing was too weak to break through an energy barrier. On the
contrary, if $\ket{\IS^*}\neq\ket{\vec{s}^{\,(\IS)}}$ the forcing was
large enough for a hop to another valley.  

To ensure well-defined forcings along $\ket{\vec\pi_\RAND}$, we
normalize $A_\F$ with $\normauno{\ket{\vec\pi}}=\sum_\bx
\left|\vec\pi_\bx\right|$. Indeed, the perturbation in~\eqref{eq:ham-forcing}
is bounded by
$\left|\sum_\bx
\vec\pi_\bx\cdot\vec{s}_\bx\right| \leq \left|\sum_\bx
\vec\pi_\bx\cdot\vec{s}_\bx\right|\leq \normauno{\ket{\vec\pi}}$.
So, the perturbation is made extensive by choosing $A_\F = {\cal A}
i_h$, with ${\cal A}=\frac{NA}{\normauno{\ket{\pi}}}$, where
the amplitudes $A$ are an external parameter (of order 1), that we
tuned in order to be in the linear response regime for small $i_h$,
and just out of it for $i_h$ approaching $N_\F$ (see appendix).

For the softest mode we analyzed the effect of forcings of order
$O(1)$ because larger ones lead the system out of the linear response
regime, so the amplitude of the forcings along $\ket{\pi_0}$ is $A_\F
(i_h) = \frac{A i_h}{\normauno{\ket{\pi}}}$.

See the appendices for further details about the
linear-response regime, hops between valleys and the phenomenology of
these rearrangements.

\begin{figure}[!t]
 \includegraphics[width=\columnwidth]{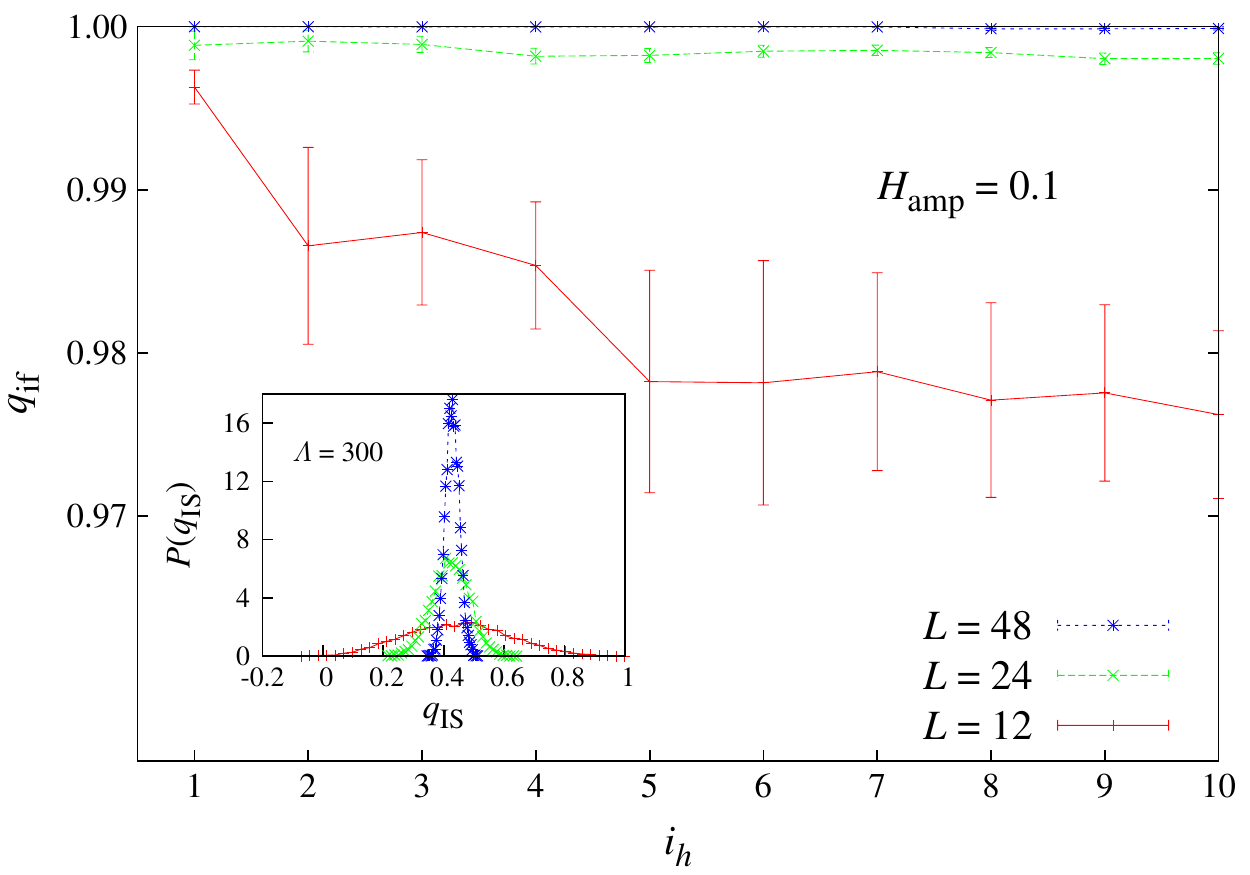}
 \caption{Overlap $q_\mathrm{if}$ between initial and forced IS (if another IS is reached) as a function of the forcing 
 step $i_h$ for $H_\amp=0.1$, for forcing along the softest mode. The inset shows the 
 distribution of the overlaps of randomly found ISs.}
 \label{fig:overlap-2_inset}
\end{figure}

\begin{figure}[!htb]
 \includegraphics[width=\columnwidth]{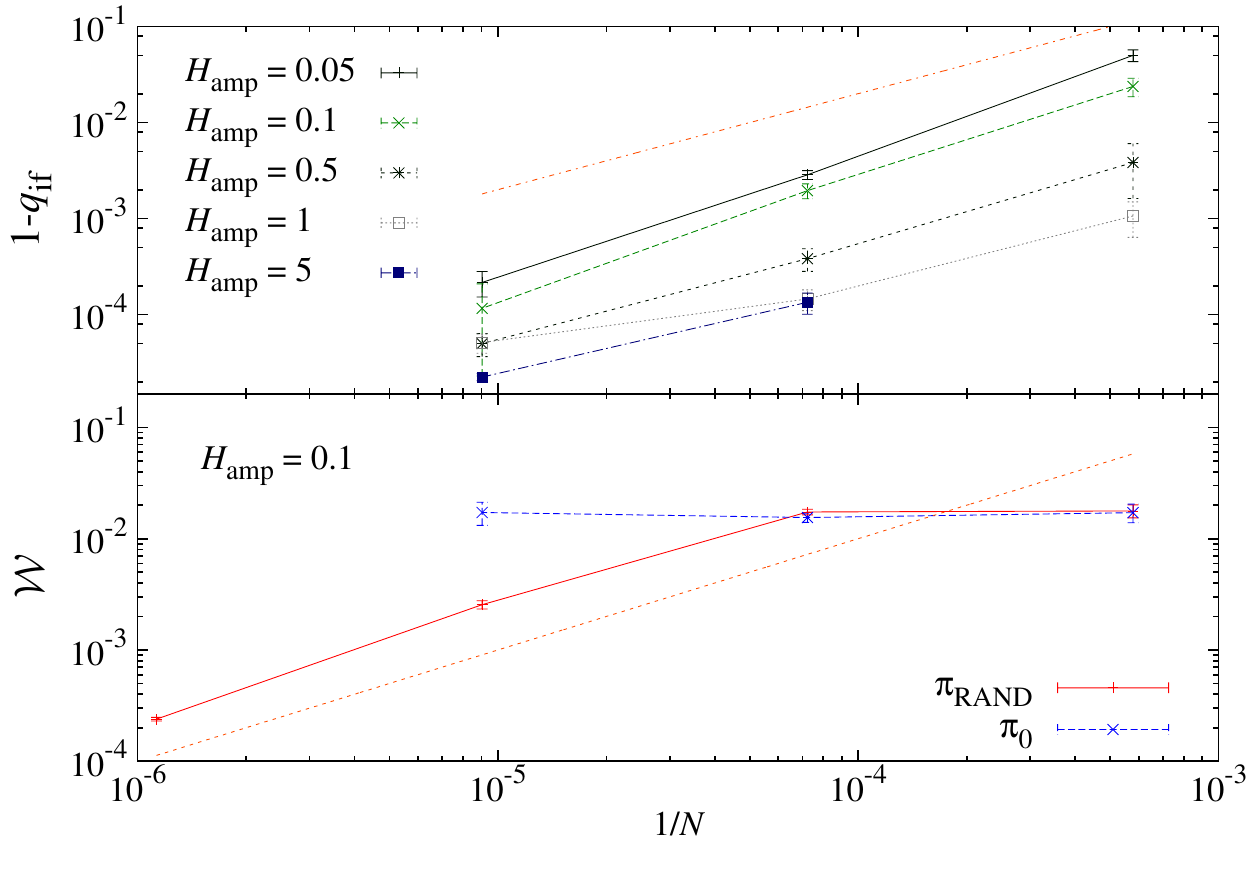}
 \caption{Top: $1-q_\mathrm{if}$ for rearrangements that occurred at the $10^\mathrm{th}$ forcing step, for fields $H_\amp=0.05,0.1,0.5,1,5$.
	  Bottom: cumulant $\mathcal{W}$ as a function of $1/N$ for $\ket{\vec\pi_\RAND}$ and $\ket{\vec\pi_0}$. In both plots, the straight line is a reference
	  curve $\propto1/N$.}
 \label{fig:overlap-1}
 \end{figure}

\paragraph{Two-level systems}
In the spectrum of $\M$, $\rho(\lambda)$, there is an extensive number of very soft modes, with a localized eigenstate. The eigenstates can connect
different ISs through the forcing procedure we described. The connection caused by such states is privileged, because the 
couples of ISs are very similar one to the other. We show this in figure \ref{fig:overlap-2_inset}, where we compare the overlap $q_\mathrm{if}$ between the configurations obtained
through a forcing of amplitude $A_\F(i_h)$ with the typical overlap between independent ISs.
\footnote{The overlap $q_\mathrm{if}$ between $\ket{\vec{s}^{\,(\IS)}}$ and $\ket{\IS(i_h)}$ is defined as $q_\mathrm{if}\equiv \frac{1}{N}\sum_\bx q_{\mathrm{if},\bx}$, 
with $q_{\mathrm{if},\bx}=\vec s_\bx^{(\IS)}\cdot \vec s_\bx(i_h)$, being $\vec s_\bx(i_h)$ the spins of the configuration $\ket{\IS(i_h)}$.}
This happens for every field that produces rearrangements (at $H_\amp=10,50$ the energy landscape is too trivial and the forcings never lead to a new IS),
as it can be seen from Fig.~\ref{fig:overlap-1}, top, where we show only the overlap $q_\mathrm{if}$ with the largest forcings, of $i_h=10$. We plot $1-q_\mathrm{if}$
and put it on log-scale so it is better visible.
The overlaps $q_\mathrm{if}$ are much closer to 1 than the overlaps of independent ISs (Fig.~\ref{fig:overlap-2_inset}, inset).
This means that the ISs are somewhat clustered in tiny
groups that are represented by a single IS. This could be an operational definition of classical TLS, 
i.e. a system in which there are two very close states, 
where the transitions from one state to the other can be treated as independent of the rest 
of the system \cite{anderson:72,phillips:72,phillips:87,lisenfeld:15,perez-castaneda:15}.

Moreover, the energy barriers separating these privileged states are positive, but they do not grow with the system size (see figure \ref{fig:energy} in 
the appendix). This suggests that in the thermodynamic limit $\ket{\IS^*}$ and $\ket{\vec{s}^{\,(\IS)}}$
are separated by an infinitely small energy barrier, just as in a TLS.

We can get more insight on the type of rearrangement that took place during the valley change, by defining the cumulant 
$\mathcal{W} = \frac{\sum_\bx^N w_\bx^2}{\left(\sum_\bx^N w_\bx\right)^2}$,
where $w_\bx=1-q_{\mathrm{if},\bx}$. If the rearrangement is completely localized, $\mathcal{W}=1$, 
whereas if it is maximally delocalized $\mathcal{W}=1/N$. 
Fig~\ref{fig:overlap-1}, bottom, shows that, as expectable, the rearrangements are localized when we stimulate the system along the softest mode, 
and delocalized when it is along a random direction.

\paragraph{Conclusions}
The introduction of a random field in the Heisenberg spin glass model, besides extinguishing the rotational symmetry, 
changes qualitatively the shape of its DOS.
Very strong random fields suppress the soft modes, and a gap appears in the DOS $g(\omega)$. Still, soft modes do resist the
application of a random field when it is not too large. The data are compatible with the absence of a gap, where for small $\omega$ the DOS
grows as $g(\omega)\propto\omega^{4}$, differently from the zero-field expectation, $g(\omega)\propto\omega^2$ \cite{grigera:11,franz:15b}.

It appears that a finite fraction of the modes is localized, suggesting a 
localization transition when the system becomes large.

We also analyzed the anharmonicity of the energy landscape, by imposing an external force on the system.
The reaction of the spin glass has a strong dependency on the direction of application of the force. 
Forcings along a random direction, need to be extensively strong in order to move the orientation of the
spins. Equivalent results, instead, can be obtained through forcings of order 1, if they are oriented along
the softest mode.

Forcings along the softest mode cause localized rearrangements that lead the system to a new IS that is infinitely similar to the original one.
The two states are separated by very small energy barriers. This could be used as
an operational definition of classical TLSs.
\nocite{baityjesi:11,baityjesi:15,amit:05,fernandez:09b,gellman:60}

\begin{acknowledgments}
We were supported by the European Research Council under the European
Union’s Seventh Framework Programme (FP7/2007-2013, ERC grant agreement 
no. 247328). 
We were partly supported by MINECO, Spain,
through the research contract N$^{o}$. FIS2012-35719-C02.
This work was partially supported by the GDRE 224 CNRS-INdAM GREFI-MEFI.
M.B.-J. was supported by the FPU program (Ministerio de Educaci\'on, Spain).
The authors thankfully acknowledge the resources from the supercomputer ``Memento'', 
technical expertise and assistance provided by BIFI-ZCAM (Universidad de Zaragoza).
\end{acknowledgments}

\appendix
\section{Appendix}

\section{Parameters of the simulations}
In Tab.~\ref{tab:hsgrf-sim} we resume how many samples we simulated for each couple $(L,H)$ both for the spectrum $\rho(\lambda)$,
and for the forcings, the number of eigenvalues $n_\lambda$ that was computed with Arpack, and the amplitudes $A$ of the forcings.
\begin{table}[!tbh]
 \centering
 \begin{tabular}{ccccccc}
  $H_\amp$ & $L$ & $n_\mathrm{samples}$ & $n_\mathrm{replicas}$ &  $n_\lambda$ &  $A (\ket{\vec\pi_0})$ &  $A (\ket{\vec\pi_\RAND})$ \\
\hline\hline
    50     & 192 &          10 (0)      &            2          &  35  &       -       &    -     \\
    50     &  96 &          10 (10)     &            2          &  80  &       -       &    1     \\
    50     &  48 &          70 (70)     &            2          & 500  &       1       &    1     \\
    50     &  24 &         100 (100)    &            2          & 500  &       1       &    1     \\
    50     &  12 &         100 (100)    &            2          & 500  &       1       &    1     \\    
\hline
    10     & 192 &          10 (0)      &            2          &  35  &       -       &    -     \\
    10     &  96 &          10 (10)     &            2          &  80  &       -       &    0.72  \\
    10     &  48 &          70 (70)     &            2          & 500  &       0.6     &    0.72  \\
    10     &  24 &         100 (100)    &            2          & 500  &       0.3     &    0.72  \\
    10     &  12 &         100 (100)    &            2          & 500  &       0.3     &    0.72  \\    
\hline
     5     & 192 &          10 (0)      &            2          &  35  &       -       &     -    \\
     5     &  96 &          10 (10)     &            2          &  80  &       -       &    0.3   \\
     5     &  48 &          70 (70)     &            2          & 500  &       0.014   &    0.3   \\
     5     &  24 &         100 (100)    &            2          & 500  &       0.02    &    0.3   \\
     5     &  12 &         100 (100)    &            2          & 500  &       0.024   &    0.3   \\    
\hline
     1     & 192 &          10 (0)      &            2          &  35  &       -       &    -     \\
     1     &  96 &          10 (10)     &            2          &  80  &       -       &    0.05  \\
     1     &  48 &          70 (70)     &            2          & 500  &       0.004   &    0.05  \\
     1     &  24 &         100 (100)    &            2          & 500  &       0.0045  &    0.05  \\
     1     &  12 &         100 (100)    &            2          & 500  &       0.0045  &    0.05  \\    
\hline
    0.5    & 192 &          10 (0)      &            2          &  35  &       -       &    -     \\
    0.5    &  96 &          10 (10)     &            2          &  80  &       -       &    0.022 \\
    0.5    &  48 &          70 (70)     &            2          & 500  &       0.008   &    0.02  \\
    0.5    &  24 &         100 (100)    &            2          & 500  &       0.009   &    0.022 \\
    0.5    &  12 &         100 (100)    &            2          & 500  &       0.009   &    0.022 \\    
\hline
    0.1    & 192 &          10 (0)      &            2          &  35  &       -       &     -    \\
    0.1    &  96 &          10 (10)     &            2          &  80  &       -       &    0.012 \\
    0.1    &  48 &         100 (70)     &            2          & 500  &       0.006   &    0.012 \\
    0.1    &  24 &         100 (100)    &            2          & 500  &       0.1     &    0.012 \\
    0.1    &  12 &         100 (100)    &            2          & 500  &       0.1     &    0.012 \\    
\hline
    0.05   & 192 &          10 (0)      &            2          &  35  &       -       &     -    \\
    0.05   &  96 &          10 (10)     &            2          &  80  &       -       &    0.011 \\
    0.05   &  48 &         100 (70)     &            2          & 500  &       0.06    &    0.011 \\
    0.05   &  24 &         100 (100)    &            2          & 500  &       0.42    &    0.011 \\
    0.05   &  12 &         100 (100)    &            2          & 500  &       0.36    &    0.011 \\    
\hline
    0.01   & 192 &          10 (0)      &            2          &  25  &       -       &      -   \\
    0.01   &  96 &          10 (10)     &            2          &  80  &       -       &    0.016 \\
    0.01   &  48 &         100 (70)     &            2          & 500  &       0.045   &    0.016 \\
    0.01   &  24 &         100 (100)    &            2          & 500  &       0.009   &    0.004 \\
    0.01   &  12 &         100 (100)    &            2          & 500  &       0.007   &    0.001 \\    
\end{tabular}
\caption[Simulation parameters from \cite{baityjesi:15b}]{Samples and replicas of our simulations. The number between parenthesis is the amount of samples
used for the forcings. We indicate with $n_\lambda$ the number of eigenvalues we calculated from the bottom of the spectrum $\rho(\lambda)$.
$A (\ket{\vec\pi_\RAND})$ and $A (\ket{\vec\pi_0})$ are the forcings' parameters.}
\label{tab:hsgrf-sim}
\end{table}

\section{Reaching the inherent structure}
As energy minimization algorithm we use the successive overrelaxation, that was successfully used in \cite{baityjesi:11} for $3d$ Heisenberg spin glasses.
It consists in an interpolation, through a parameter $\Lambda$, between a direct quench, that aligns all the spins to the local field $\vec h_\bx$ (see. e.g. \cite{baityjesi:15} for a recent 
application in Heisenberg spin glasses), and the microcanonical overrelaxation update (well explained in \cite{amit:05}).

We propose sequential single-flip updates with the rule
\begin{equation}
 \vec{s}_\bx^{\,\mathrm{SOR}} = \frac{\vec{h}_\bx + \Lambda \vec{s}_\bx^\mathrm{\,OR}}{||\vec{h}_\bx + \Lambda \vec{s}_\bx^\mathrm{\,OR} ||}\,,
\end{equation}
where $\vec{s}_\bx^\mathrm{\,OR}$ is the overrelaxation update
\begin{equation}
  \vec s_{\bx}^\mathrm{\,OR}=2\vec h_{\bx}\frac{\vec h_{\bx}\cdot\vec
    s_{\bx}^\mathrm{\,old}}{h_{\bx}^2} - \vec s_{\bx}^\mathrm{\;old}\,.
\end{equation}
The limit $\Lambda=0$ corresponds to a direct quench that notoriously presents convergence problems. On
the other side, with $\Lambda=\infty$ the energy does not decrease.

It is shown in \cite{baityjesi:11} that the optimal value of $\Lambda$ in terms of
convergence speed is $\Lambda\approx300$.
Thus, the search of ISs was done with $\Lambda = 300$, under the reasonable
assumption, that we will reinforce right away, that a change on $\Lambda$ does not imply sensible changes 
in the observables we examine.
In fact, the concept of IS is strictly related to the protocol one chooses to relax the system, and on the starting
configuration. From \cite{baityjesi:11} our intuition is that despite the ISs' energies do depend on these two 
elements, this dependency is small and we can neglect it (dependencies on the correlation lengths will be examined in a future work \cite{baityjesi:16}).

To validate the generality of our results we compared the ISs reached with $\Lambda=300, 1, 0$, at $H_\amp=0$ over a wide range of temperatures.
We took advantage, for this comparison, of the $L=48$ configurations that were thermalized in \cite{fernandez:09b}, that go from
$T_\SG$ to $\frac{5}{3}T_\SG$.

\begin{figure}[!htb]
 \includegraphics[width=\columnwidth]{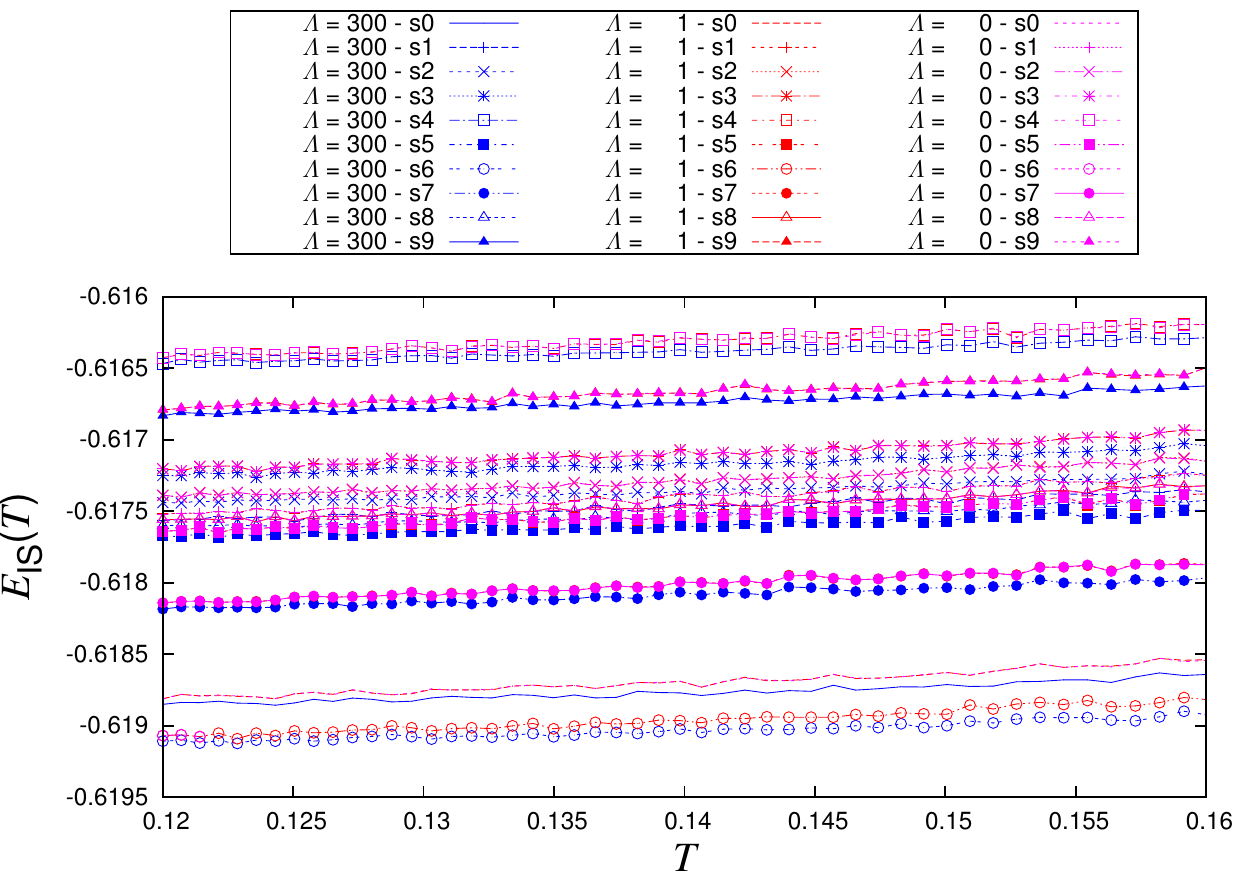}
  \caption{
  Energy of the inherent structure as a function of temperature for 10 samples chosen at random, for $H_\amp=0$, $L=48$.
  We use the same symbol for the same sample. ISs obtained with $\Lambda=300$ are in blue, red represents $\Lambda=1$ and purple
  stands for $\Lambda=0$. Purple and red lines almost overlap.
  Sample-to-sample fluctuations are the largest source of dispersion, compared with $\Lambda$ and $T$.}  
  \label{fig:ecompare-lambda}
\end{figure}

In figure \ref{fig:ecompare-lambda} we plot the energy $E_\IS$ of the reached ISs,
\footnote{With a normalization factor $1/3N$ that bounds it to unity.}
as a function of the temperature $T$.
We show ten random samples, each minimized with $\Lambda=300,1,0$. 
Increasing $\Lambda$ the energy of the inherent structures decreases but this variation is smaller than the 
dispersion between different samples. The energy of the ISs also decreases with $T$, but this decrease too is smaller than the
fluctuation between samples.
Since the dispersion on the energy is dominated by the disorder, rather than by $\Lambda$ or $T$, we can think of putting ourselves in the most
convenient situation: $T=\infty$, that does not require thermalization, and $\Lambda=300$, that yields the fastest minimization.

\begin{figure}[ht]
 \includegraphics[width=\columnwidth]{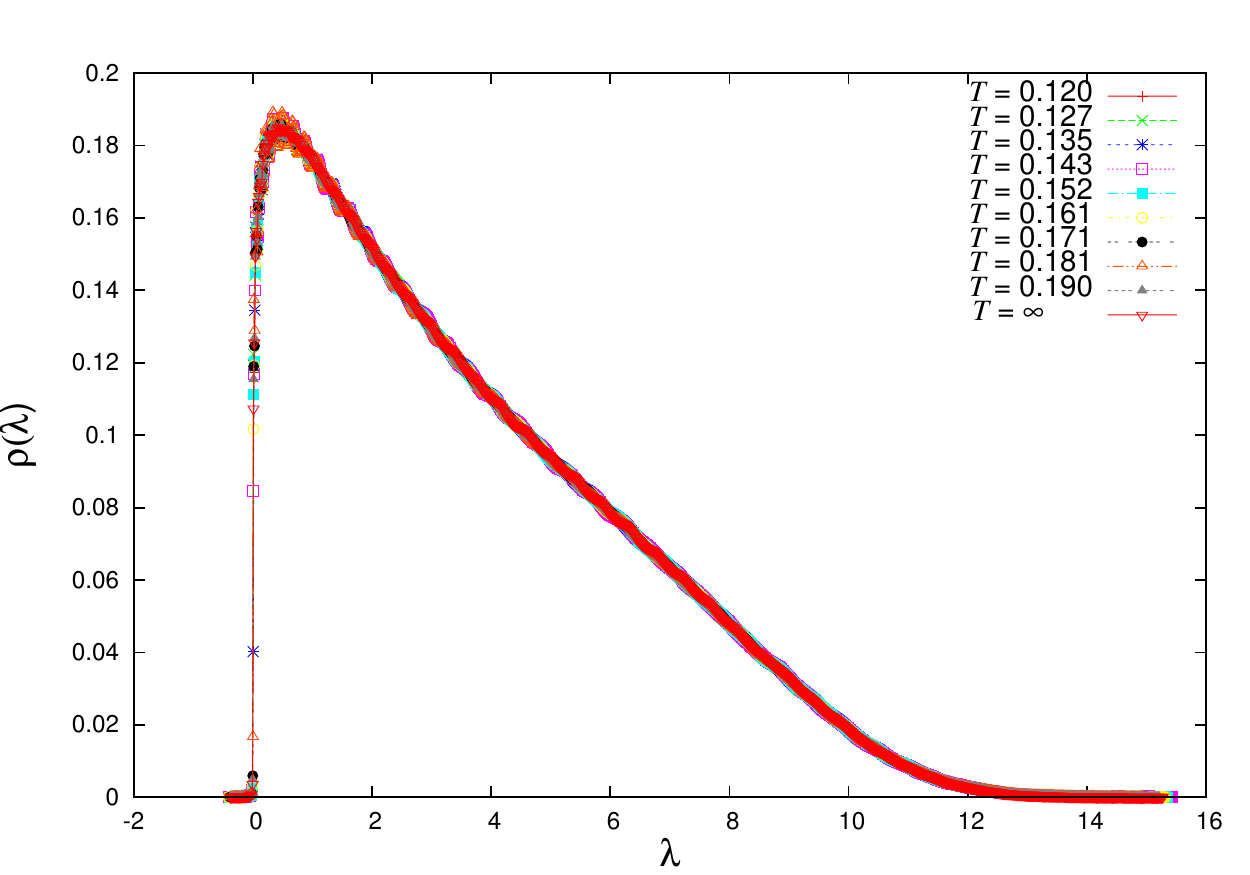}
 \includegraphics[width=\columnwidth]{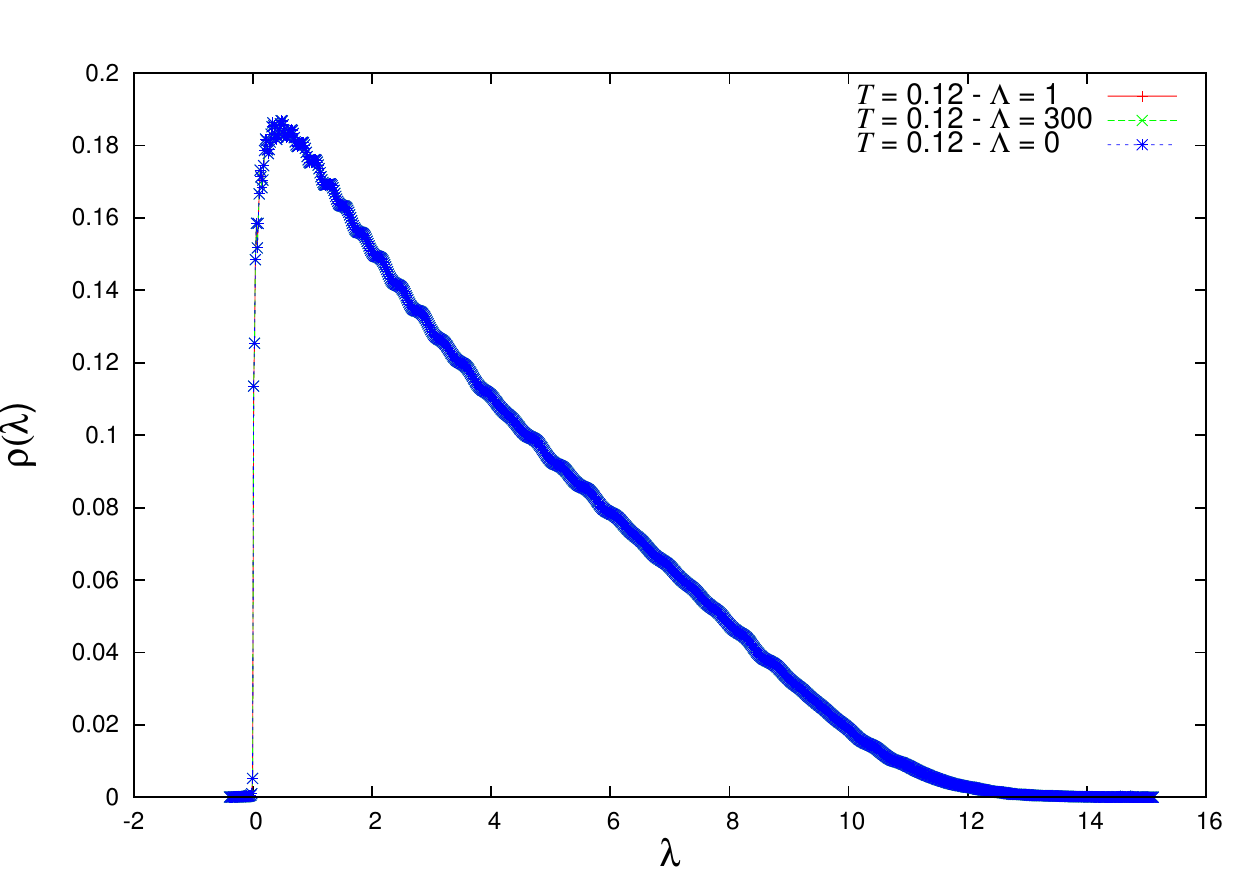}
 \caption[Dependency of the spectrum on $T$ and $\Lambda$.]{Spectrum $\rho(\lambda)$ of the Hessian matrix 
 calculated at the inherent structure for $H_\amp=0$, $L=48$, with the method of the moments .
 \textbf{Top}: $\rho(\lambda)$ for different temperatures from $T=0.12$ to $T=\infty$. 
 \textbf{Bottom}: comparison of the spectrum between $\Lambda=1$ and $\lambda=300$ at $T=0.12, 0.19, \infty$ various $\Lambda$.
 }
  \label{fig:spectrum-dep}
\end{figure}

Also the spectrum of the dynamical matrix does not show relevant signs of dependency on either $T$ of $\Lambda$, as shown in figure \ref{fig:spectrum-dep}.

\section{The Dynamical matrix}
We calculate the dynamical matrix as the Hessian matrix $\M$ of Hamiltonian 
\begin{equation}\label{eq:HRFapp}
 \mathcal{H}_\RF(\ket{\vec s}) = -\sum_{|\bx - \by|=1} J_{\bx\by} \vec{s}_{\bx}\cdot \vec{s}_{\by} - \sum_{\bx}^{N} \vec{h}_{\bx} \cdot \vec{s}_{\bx}\,,
\end{equation}
calculated at the local minima of the energy.
It is not straightforward to compute $\M$ because it is necessary to take in account the normalization of the spins  $\vec{s}_\bx^{\,2}=1~\forall\bx$.

To this scope we define local perturbation vectors $\vec{\pi}_\bx$, called pions in analogy with the nonlinear $\sigma$ model \cite{gellman:60}.
The distinguishing feature of the pions is that they are orthogonal to the IS, $\left(\vec{s}_\bx\cdot \vec{\pi}_\bx\right)=0$, and that their global
norm is unitary, $\langle\vec{\pi}\ket{\vec{\pi}}=1$, where $\langle\vec a\ket{\vec b}$ indicates the scalar product between configurations, 
$\langle \vec a\ket{\vec b} \equiv \sum_\bx^N \vec a_\bx\cdot \vec b_\bx$.

We use the pions to parametrize order $\epsilon$ perturbations around the IS as
\begin{equation}\label{eq:pions}
 \vec{s}^{\,\epsilon}_\bx = \vec{s}_\bx^\mathrm{\,(IS)} \sqrt{1-\epsilon^2\vec{\pi}_\bx^2} + \epsilon\vec{\pi}_\bx\,,
\end{equation}
so the position of $\vec{s}^{\,\epsilon}_\bx$ is fully determined by $\vec{\pi}_\bx$.
As long as $\epsilon$ is small enough to grant $\epsilon^2\vec{\pi}_\bx^2<1$ $\forall\bx$, the normalization condition is naturally satisfied without the need 
to impose any external constraint.

We now build a local reference change. For each site $\bx$ we define a local basis $\mathcal{B}=\left\{\vec{s}_\bx^\mathrm{\,(IS)}, \hat{e}_{1,\bx}, \hat{e}_{2,\bx}\right\}$,
where $\hat{e}_{1,\bx}, \hat{e}_{2,\bx}$ are any two unitary vectors, orthogonal to each other and to $\vec{s}_\bx^\mathrm{\,(IS)}$, and well oriented. In our simulations they were generated
randomly. In this basis the pions can be rewritten as $\vec{\pi}_\bx = (0, a_1, a_2)$, where now they explicitly depend only on two components, with 
real values $a_1$ and $a_2$. We can therefore rewrite the pions as two-component vectors $\tilde{\pi}_\bx = (a_1, a_2)$.
At this point we integrated the normalization constraint with the parametrization, and we can obtain the $2N\times2N$ Hessian matrix $\M$, that acts on
$2N$-component vectors $\ket{\tilde{\pi}}$, by a second-order development of $\mathcal{H}_\IS$ (the derivation of $\M$ is shown in the subsection ahead). 
The obtain matrix is sparse, with 13 non-zero elements per line (1 diagonal element, and 6 two-component vectors for the nearest-neighbors). The matrix element $\M^{\alpha\beta}_{\bx\by}$ is 
\begin{equation}
 \M^{\alpha\beta}_{\bx\by} = \M_{\bx\by} (\hbe_{\alpha,\bx}\cdot\hbe_{\beta,\by})\,,\\
 \end{equation}
 with
 \begin{equation}
 \M_{\bx\by} = \delta_{\bx\by} (\vec{h}_\by^\mathrm{\,(IS)}\cdot\vec{s}_\by^\mathrm{\,(IS)})- \sum_{\mu=-D}^{D}J_{\bx\by}\delta_{\bx+\hat\mu, \by}\,,
\end{equation}
where the bold arab characters indicate the site, and the greek characters indicate the component of the two-dimensional vector.

\subsection{Derivation of the expression for the Hessian \label{app:hsgrf-hessian}}
We derive the expression of the Hessian matrix $\M$ of the Hamiltonian $\mathcal{H}_\RF$ [eq. (\ref{eq:HRFapp} in the main article] that we implemented in our programs.

In terms of pionic perturbations [recall eq. (\ref{eq:pions})], $\M$ would be defined as 
$\M_{\bx\by}^{\alpha\beta}=\frac{\partial^2\mathcal{H}_\RF}{\partial\pi_{\bx,\alpha}\pi_{\by,\beta}}$.
An easy way to extract the Hessian is to write $\mathcal{H}_\RF$ as perturbations around the IS and to pick only the second-order terms.

To rewrite $\mathcal{H}_\RF$ as a function of the pionic perturbations, it is simpler to compute separately the dot products $\left(\vec{s}_\bx\cdot \vec{s}_\by\right)$
and $\vec{h}_{\bx} \cdot \vec{s}_{\bx}$.
Including the $\epsilon$ factors into the perturbation $\pi_x$, the generic spin near the IS is expressed as 
$\vec{s}_\bx = \vec{s}^{\,(\IS)}_\bx\sqrt{1-\vec{\pi}_\bx^2} + \vec{\pi}_\bx$.
We can make a second-order expansion of the non-diagonal part of the Hamiltonian by taking
the first-order expansion of the square root $\sqrt{1-\vec{\pi}_\bx^2}\simeq1-\vec{\pi}_\bx^2/2$,
\begin{align}\label{eq:term-coupling}
 &\left(\vec{s}_\bx\cdot \vec{s}_\by\right) =\\
=&   \left(\vec{s}^{\,(\IS)}_\bx\sqrt{1-\vec{\pi}_\bx^2} + \vec{\pi}_\bx\right) \cdot\left(\vec{s}^{(\,\IS)}_\by\sqrt{1-\vec{\pi}_\by^2} + \vec{\pi}_\by\right) =\nonumber
\end{align}
\vspace{-.5cm}
\begin{multline}
 =\left(\vec{s}^{\,(\IS)}_\bx\cdot\vec{s}^{\,(\IS)}_\by\right)+\left(\vec{s}^{\,(\IS)}_\bx\cdot\vec{\pi}_\by\right)+
\left(\vec{s}^{\,(\IS)}_\by\cdot\vec{\pi}_\bx\right)+\\
+\frac{1}{2}\left[\left(-\vec{\pi}_\bx^2-\vec{\pi}_\by^2\right)\left(\vec{s}^{\,(\IS)}_\bx\cdot\vec{s}^{\,(\IS)}_\by\right)+2\vec{\pi}_\bx\cdot\vec{\pi}_\by\right]+o(\vec\pi^3)\,.\nonumber
\end{multline}
On the other hand the random-field term is
\begin{equation}\label{eq:term-rf}
\left(\vec{h}_\bx\cdot \vec{s}_\bx\right) =\vec{h}_\bx\cdot\left(\vec{s}^{\,(\IS)}_\bx\sqrt{1-\vec{\pi}_\bx^2} + \vec{\pi}_\bx\right) =
\end{equation}
\vspace{-.5cm}  
\begin{equation}
=  \left(\vec{h}_\bx\cdot\vec{s}^{\,(\IS)}_\bx\right) + \left(\vec{h}_\bx\cdot\vec{\pi}_\bx\right)
  -\frac{\vec{\pi}_\bx^2}{2}\left(\vec{h}_\bx\cdot \vec{s}^{\,(\IS)}_\bx\right)+o(\vec\pi^3) \,.\nonumber
\end{equation}
 By inserting eqs.(\ref{eq:term-coupling},\ref{eq:term-rf}) and taking only the second-order terms we obtain how the 
Hessian matrix acts on the fields $\ket{\pi}$:
\begin{equation}
 \frac{1}{2}\bra{\vec\pi_\bx}\M\ket{\vec\pi_\by} =
 \end{equation}
 \vspace{-.5cm}
\begin{multline} 
 =-\frac{1}{2}\sum_{<\bx,\by>} J_{\bx,\by}\left[\left(-\vec{\pi}_\bx^2-\vec{\pi}_\by^2\right)
 \left(\vec{s}^{\,(\IS)}_\bx\cdot\vec{s}^{\,(\IS)}_\by\right)+\right.\\
 \left.+ 2\vec{\pi}_\bx\cdot\vec{\pi}_\by\right]+\sum_\bx^N \frac{\vec{\pi}_\bx^2}{2}\left(\vec{h}_\bx\cdot \vec{s}^{\,(\IS)}_\bx\right) =\nonumber
 \end{multline}
\vspace{-.5cm}
\begin{multline}  
=\frac{1}{2}\sum_\bx^N\vec{\pi}_\bx^2\left[\vec{s}^{\,(\IS)}_\bx\cdot\left(\vec{h}^{\,(\IS)}_\bx+\vec{h}_\bx\right)\right]+\\
+\frac{1}{2}\sum_\bx\vec{\pi}_\bx\cdot\sum_{\by:\norm{\bx-\by}=1}J_{\bx\by}\vec{\pi}_\by\nonumber\,,
\end{multline}
where we called $\vec{h}^{\,(\IS)}_\bx$ the local field of the IS. 
The just-obtained expression represents a sparse matrix with a matrix element $\M_{\bx\by}$ that comfortably splits as 
$\M_{\bx\by}=\mathcal{D}_{\bx\by}+\mathcal{N}_{\bx\by}$ into a diagonal
term $\mathcal{D}_{\bx\by}$ and a nearest-neighbor one $\mathcal{N}_{\bx\by}$, with
\begin{align}
 \mathcal{D}_{\bx\by} &= \delta_{\bx\by} \left[\vec{s}^{\,(\IS)}_\bx\cdot\left(\vec{h}^{\,(\IS)}_\bx+\vec{h}_\bx\right)\right]\,,\\[1ex]
 \mathcal{N}_{\bx\by} &= -\sum_{{\mu}=-d}^d J_{\bx\by} \delta_{\bx+\hat{e}_\mu,\by}\,,
\end{align}
where $\hat{e}_\mu$ is the unit vector towards one of the 2$d$ neighbors.

\paragraph{$\M$ in the local reference frame} 
The last step is to get an expression of the Hessian matrix in the local reference frame, that includes the spin normalization constraint.

In the local reference frame the pions are written like $\vec{\pi}=a_1\hat{e}_{1,\bx}+ a_2\hat{e}_{2,\bx}$ because they are perpendicular to the first
vector of the basis, $\vec{s}^{\,(\IS)}_\bx$, and that is why we write them in a two-dimensional representation as 
$\tilde{\pi}=(a_1,a_2)$.

In this local basis, the matrix element acting on the pions is written as
\begin{equation}
  \vec{\pi}_\bx\M_{\bx\by}\vec{\pi}_\by =\vec{\pi}_\bx\M_{\bx\by}\vec{\pi}_\by =
\end{equation}
\vspace{-.2cm}
\footnotesize
\begin{equation*}	
 = (a_{1,\bx}, ~a_{2,\bx})\begin{pmatrix}
                                                         \M_{\bx\by}(\hat{e}_{1,\bx}\cdot\hat{e}_{1,\by})&\M_{\bx\by}(\hat{e}_{2,\bx}\cdot\hat{e}_{1,\by})\\
                                                         \M_{\bx\by}(\hat{e}_{1,\bx}\cdot\hat{e}_{2,\by})&\M_{\bx\by}(\hat{e}_{2,\bx}\cdot\hat{e}_{2,\by})
                                                        \end{pmatrix}
							\left(\begin{array}{c}
								a_{1,\by}\\a_{2,\by}
                                                              \end{array}\right)\,,
\end{equation*}
\normalsize
so in the 2$N$-dimensional reference $\M$ is expressed as
\begin{equation}
\M_{\bx\by}^{\alpha\beta}=\M_{\bx\by}\left(\hat{e}_{\alpha,\bx}\cdot\hat{e}_{\beta,\by}\right)\,, 
\end{equation}
and the elements of the diagonal and nearest-neighbor operators $\mathcal{D}$ and $\mathcal{N}$ become
\begin{align}
 \mathcal{D}_{\bx\by}^{\alpha\beta} &= \delta_{\bx\by}\,\delta^{\alpha\beta} \left[\vec{s}^{\,(\IS)}_\bx\cdot\left(\vec{h}^{\,(\IS)}_\bx+\vec{h}_\bx\right)\right]\,,\\[1ex]
 \mathcal{N}_{\bx\by}^{\alpha\beta} &= -\sum_{{\mu}=-d}^d J_{\bx\by} \delta_{\bx+\hat{e}_\mu,\by}\left(\hat{e}_{\alpha,\bx}\cdot\hat{e}_{\beta,\by}\right)\,.
\end{align}

\section{Cumulatives for all the fields}
To calculate $F(\lambda)$, for numerical purposes we compute the average of the $k^\mathrm{th}$ eigenvalue, and we plot $k/(2N)$ versus $\langle \lambda_k\rangle$. 
In this way we deliberately avoid to looking 
to the tail in $F(\lambda)$ that in finite volume systems is present as an effect of the fluctuation of the lowest eigenvalues. With these procedure the errorbars are on 
the $x$ axis. An advantage is that the function $F(\lambda)$  does not depend on the number of eigenvalue computed.

\begin{figure}[!t]
 \includegraphics[width=0.49\columnwidth]{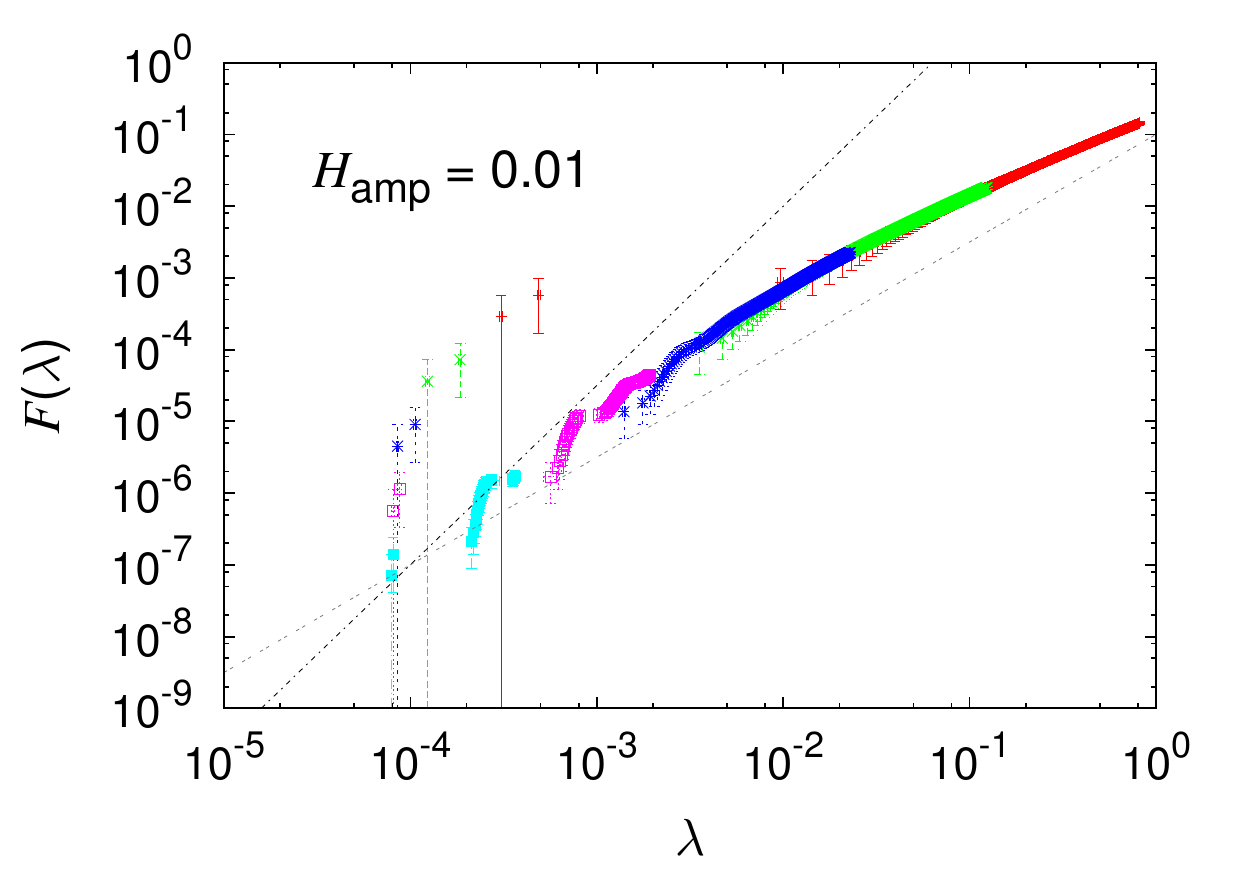}
 \includegraphics[width=0.49\columnwidth]{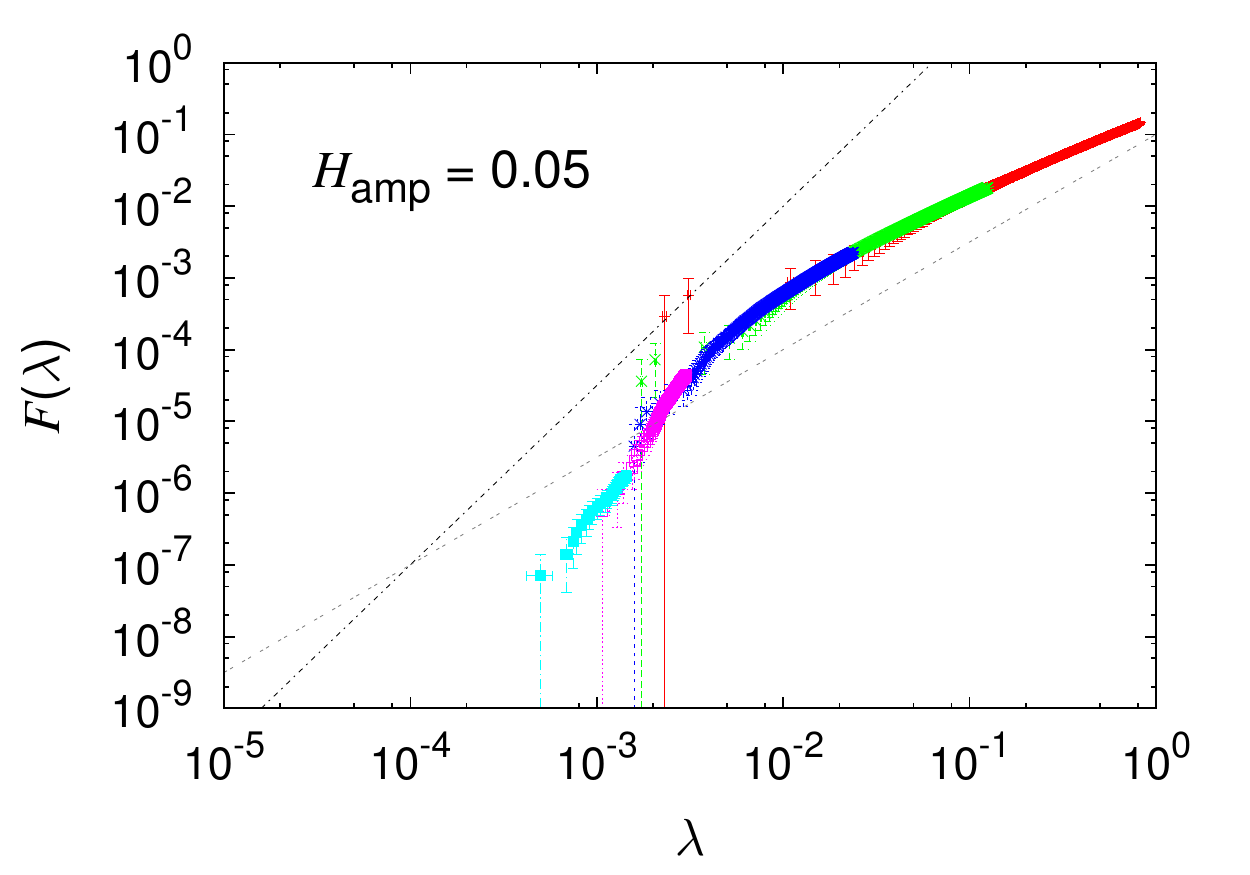}
 \includegraphics[width=0.49\columnwidth]{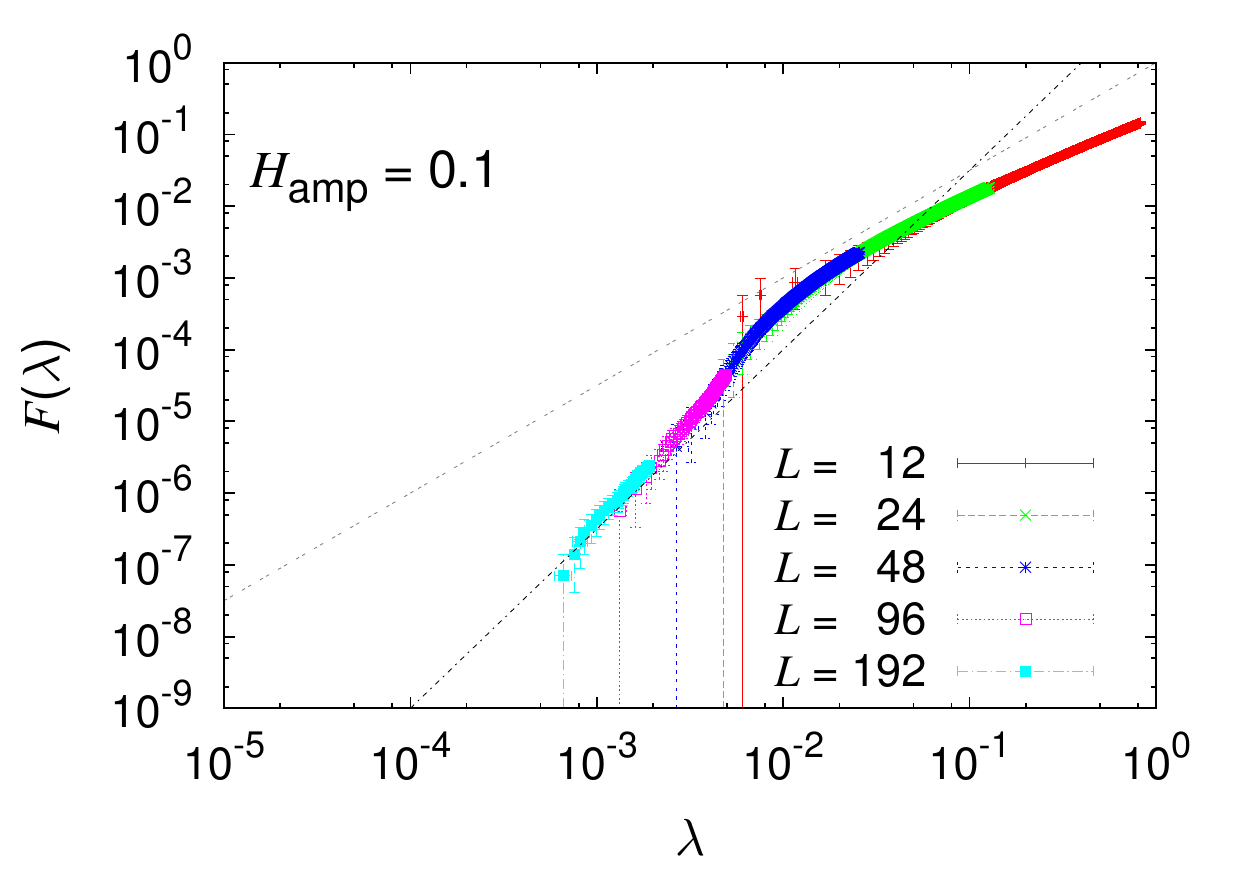}
 \includegraphics[width=0.49\columnwidth]{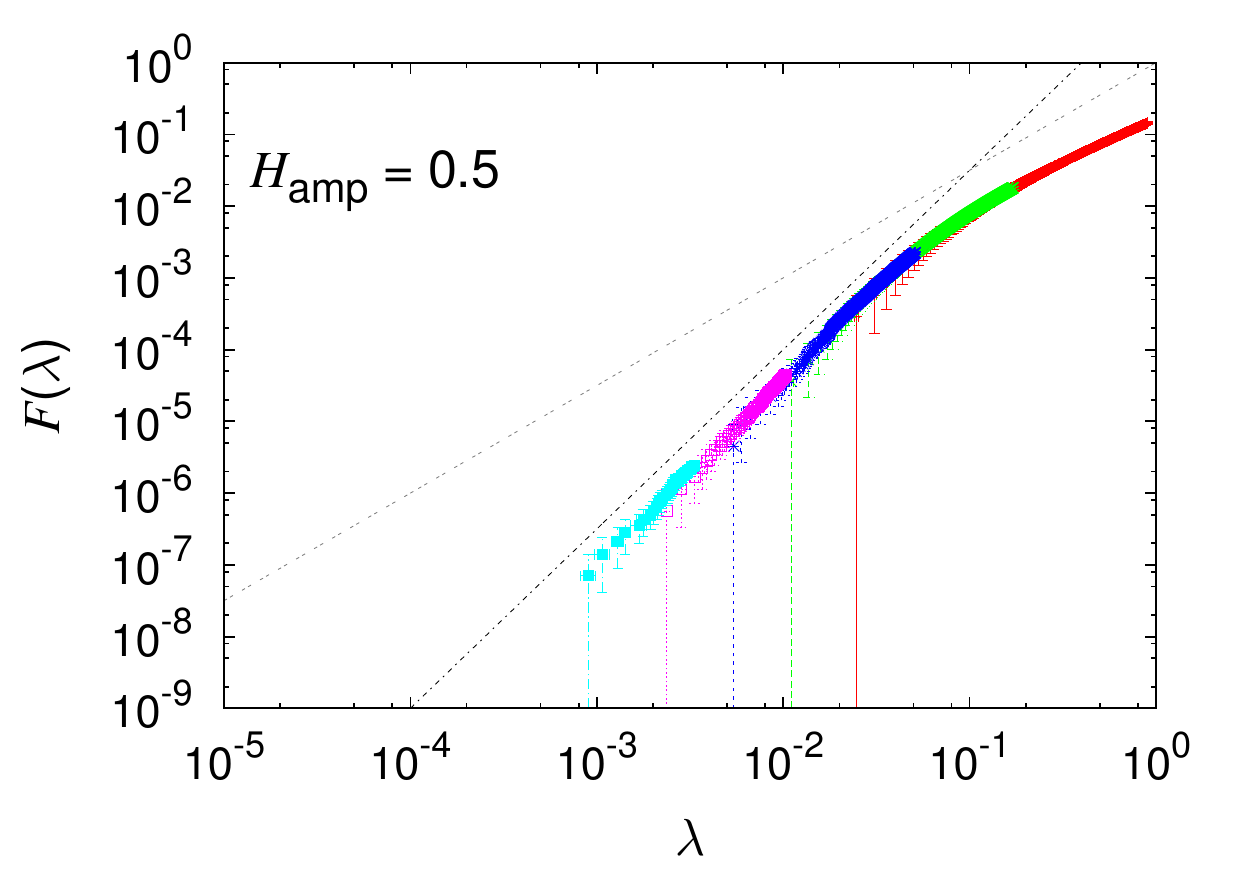}
 \caption[Cumulative distributions $F(\lambda)$ for small random fields]
	 {Cumulative distributions $F(\lambda)$ for small random fields $H_\amp = 0.01, 0.05, 0.1, 0.5$. In each plot we show black a reference curve representing the power law $\lambda^{2.5}$,
	 that is our guess for a universal behavior, and a grey line indicating the Debye behavior $\lambda^{1.5}$. One could expect a Debye behavior for $\lambda>\lambda^*$,
	 with $\lambda^{*}\to0$ as $H_\amp\to0$. Instead, we see an excess of eigenvalues even compared to the Debye behavior, indicating a likely \index{boson peak} boson peak.
	 Further discussions in the main text.} 
	 \label{fig:F-1}
\end{figure}

\begin{figure}[!t]
 \includegraphics[width=0.49\columnwidth]{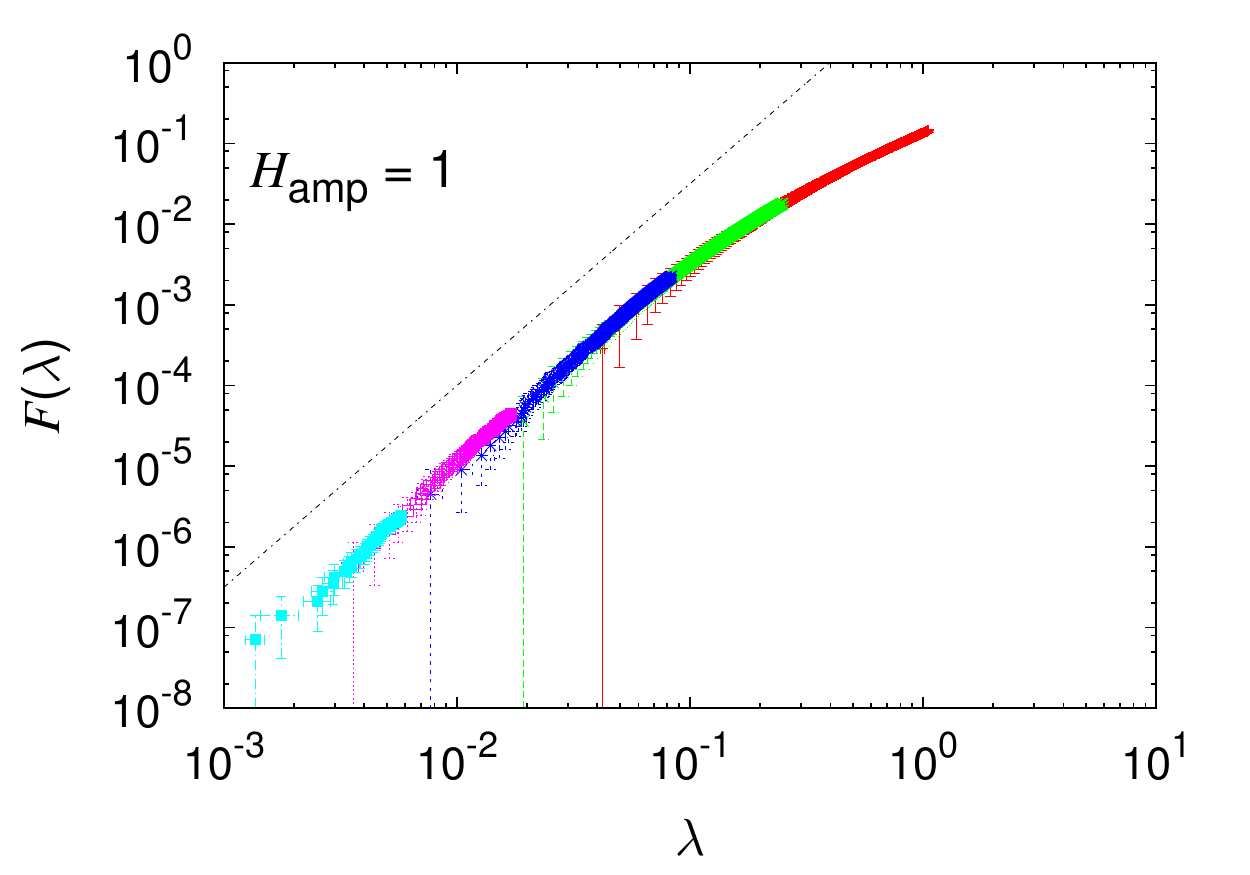}
 \includegraphics[width=0.49\columnwidth]{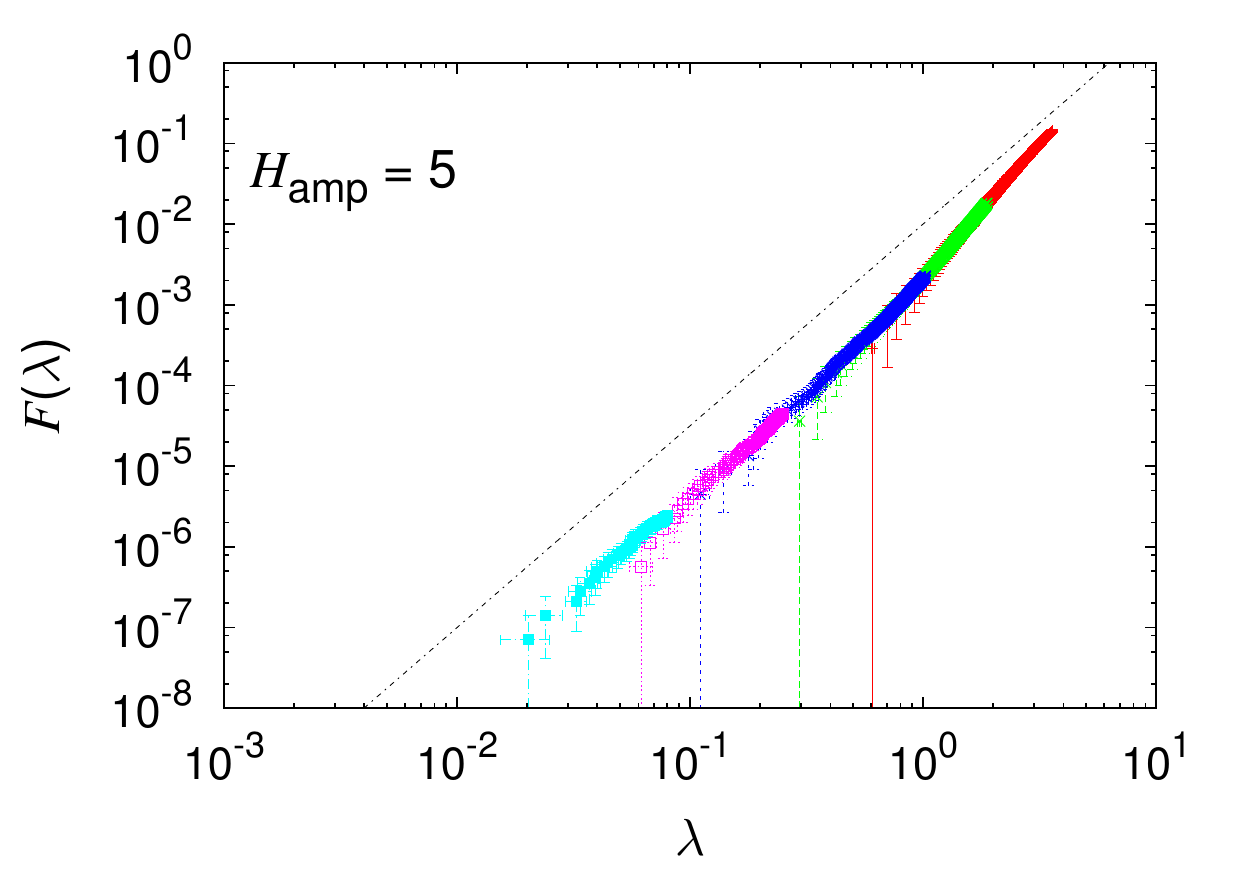}
 \includegraphics[width=0.49\columnwidth]{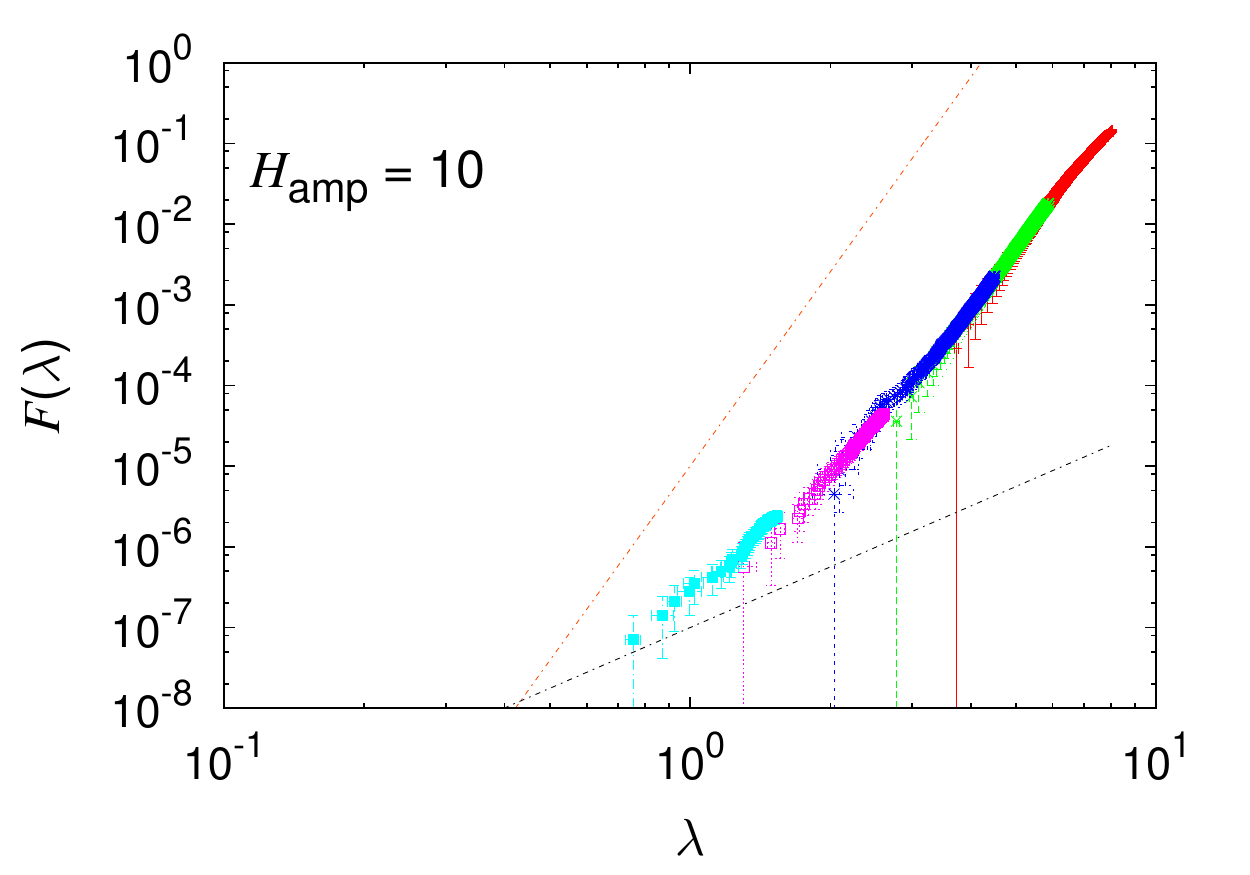}
 \includegraphics[width=0.49\columnwidth]{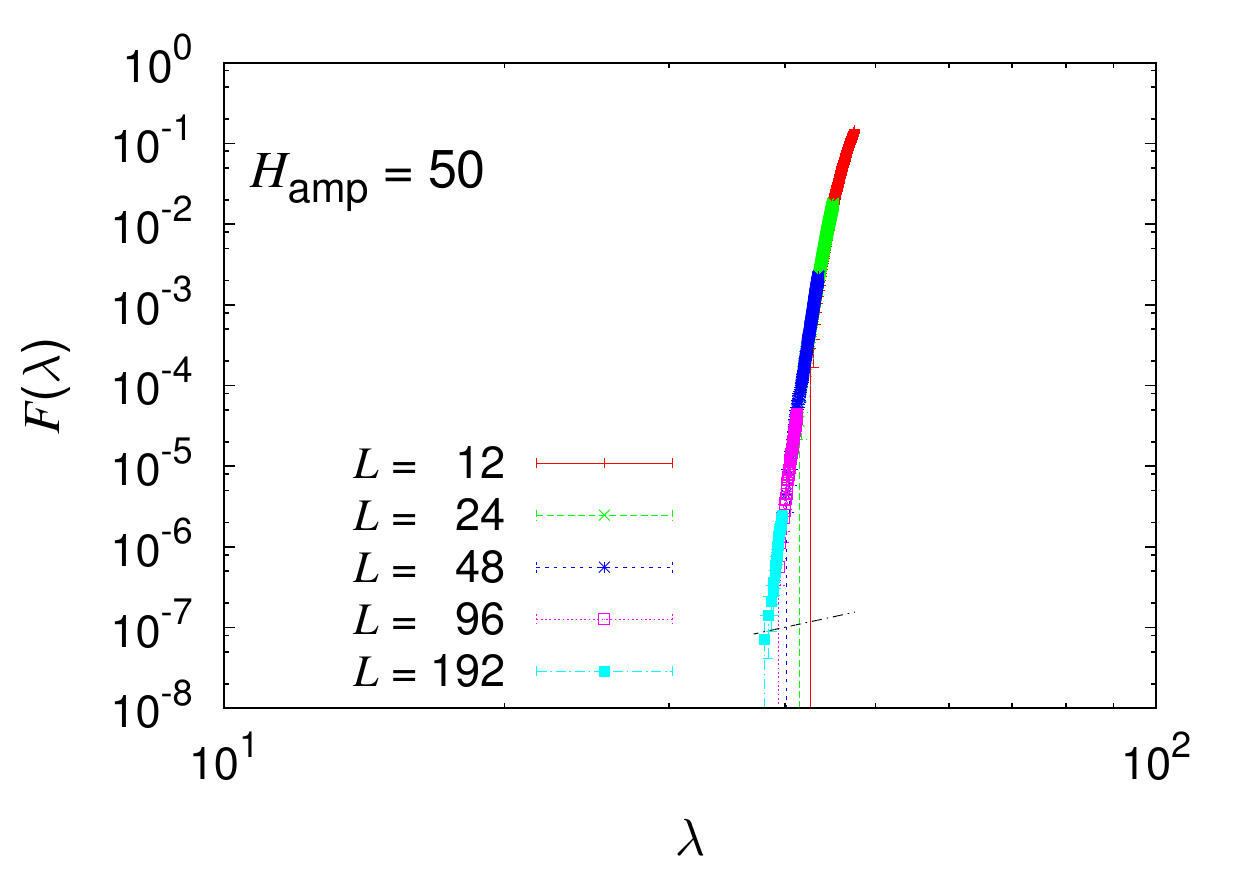}
 \caption[Cumulative distributions $F(\lambda)$ for large random fields]
         {Cumulative distributions $F(\lambda)$ for large random fields $H_\amp = 1, 5, 10, 50$.In each plot we show a reference curve representing the power law $\lambda^{2.5}$.
         The orange line in the \textbf{bottom left} set is proportional to $\lambda^8$.}
 \label{fig:F-2}
\end{figure}

In figures \ref{fig:F-1} and \ref{fig:F-2} we show the function $F(\lambda)$ for all the fields we simulated. We were able to calculate with Arpack the lowest eigenvalues 
of the spectrum. The number of calculated eigenvalues $n_\lambda$ is in table \ref{tab:hsgrf-sim}.
All the plots are compared with the Debye behavior $\lambda^{1.5}$ and with the power law behavior $\lambda^{2.5}$, 
because our data suggest that a universality on the exponents would set them around 
around $\gamma=2.5$, $\delta=4$ and $\alpha=1.5$.
This is straightforward for $H_\amp = 0.1,0.5, 1, 5$, where when $\lambda$ is small
there is a clear power law behavior, with a power close to 2.5, while it can be excluded for $H_\amp=50$, where the soft modes are suppressed in favor of a gap,
as it was also visible from the inset of figure 1 of the main article.
At $H_\amp=10$ we are probably close to where the gap forms. The $F(\lambda)$ goes as a large power law $\lambda^{8}$ when $\lambda$ is large, but at the
smallest values of $\lambda$, recovered from $L=192$, there is a slight change of power law towards something that could become 2.5.
One could also argue that a $F(\lambda)$ goes to zero as a power law for any finite $H_\amp$, as long as one looks at small enough $\lambda$. Numerical analysis cannot answer to questions of this type,
but still, even if no sharp transition is present, an empirical gap is clearly present for large $H_\amp$, since the precision of any experiment (numerical or real) is finite.
In the case of the smallest fields $H_\amp=0.01, 0.05$, we suffer from effects from $H_\amp=0$. The spin waves do not hybridize with the bulk of the spectrum, and pseudo-Goldstone
modes with a very small eigenvalue appear, making it hard to extract a power law behavior.

Overall, we see good evidence for a $\gamma$ around $2.5$ at several values of $H_\amp$, and at other fields the data is not in contradiction with a hypothesis of universality
in the exponents defined by 
\begin{equation}
  g(\omega)    \sim\omega^\delta\,,~~~
  \rho(\lambda)\sim\lambda^\alpha\,,~~~
  F(\lambda)   \sim\lambda^\gamma\,.
\end{equation}
When the field is small we remark a change of trend from $\gamma\approx2.5$ to $\gamma<1.5$ at a value $\lambda^*$. The crossover $\lambda^*$ 
shifts towards zero as $H_\amp$ decreases. This probably indicates the presence of a \index{boson peak} boson peak, an excess of modes at low frequency. Signs of a boson peak in at $H_\amp=0$ can
be seen in figure \ref{fig:spectrum-dep}. In that case the mass of the spectrum is all concentrated at low $\lambda$, but there ought to be a Debye behavior, meaning that $\lambda^*$
is very little.

\section{Eigenvector correlation function}
Since in a localized state the eigenvectors have a well-defined correlation length, we can use also this criterion to probe the localization.
We can define a correlation length from Green's function $\mathcal{G}$, that is defined through the relation $\M\mathcal{G} = \delta_{\bx\by}$,
an is commonly used in field theory for two-point correlations.
Since $\M^{-1}$ shares eigenvectors $\ket{\psi_n}$ with $\M$ and has inverse eigenvalues $1/\lambda_n$, Green's function is
\footnote{For simplicity we use $N$-component eigenvectors $\psi_n(\bx)$ instead of the $2N$-component ones $\ket{\tilde\pi}$. The relationship
between the two can be recovered through $\psi_n^2(\bx)=\tilde\pi^2=\vec\pi^2$.}
\begin{equation}
 \mathcal{G}(\bx,\by) = \M^{-1}\delta_{\bx\by} = \sum_n \frac{\psi_n(\bx)\psi_n(\by)}{\lambda_n}\,,
\end{equation}
and squaring the relation
\begin{equation}
  \mathcal{G}^2(\bx,\by) = \sum_{m,n} \frac{\psi_m(\bx)\psi_m(\by)\psi_n(\bx)\psi_n(\by)}{\lambda_m\lambda_n}\,.
\end{equation}
By averaging over the disorder we gain translational invariance and $\overline{\mathcal{G}^2}$ can be written as a function of the distance $\br=\bx-\by$,
\footnotesize
\begin{equation}
 \overline{\mathcal{G}^2(\br)} =  \overline{\sum_{m,n} \frac{1}{\lambda_m\lambda_n} \sum_\bx\left(\frac{[\psi_m(\bx)\psi_n(\bx)][\psi_m(\bx+\br)\psi_n(\bx+\br)]}{V}\right)}\,.
\end{equation}
\normalsize
Making the reasonable assumption that different eigenvectors do not interfere with each other, and exploiting the orthogonality condition $\sum_x\psi_m(\bx)\psi_n(\bx)=\delta_{mn}$,
we obtain the desired correlation function
\begin{equation}
 \mathcal{C}(\br) = \overline{\mathcal{G}^2(\br)} = \overline{\sum_{n} \frac{1}{\lambda_n^2} \psi^2_n(\bx)\psi_n^2(\bx+\br)}\,.
\end{equation}
This correlation function favors the softest modes by a factor $1/\lambda_n^2$. This is an advantage, because the bulk modes do not exhibit a finite correlation length,
so it is useful to have them suppressed.

\subsection{The correlation function for small fields}
\begin{figure}[h]
 \includegraphics[width=\columnwidth]{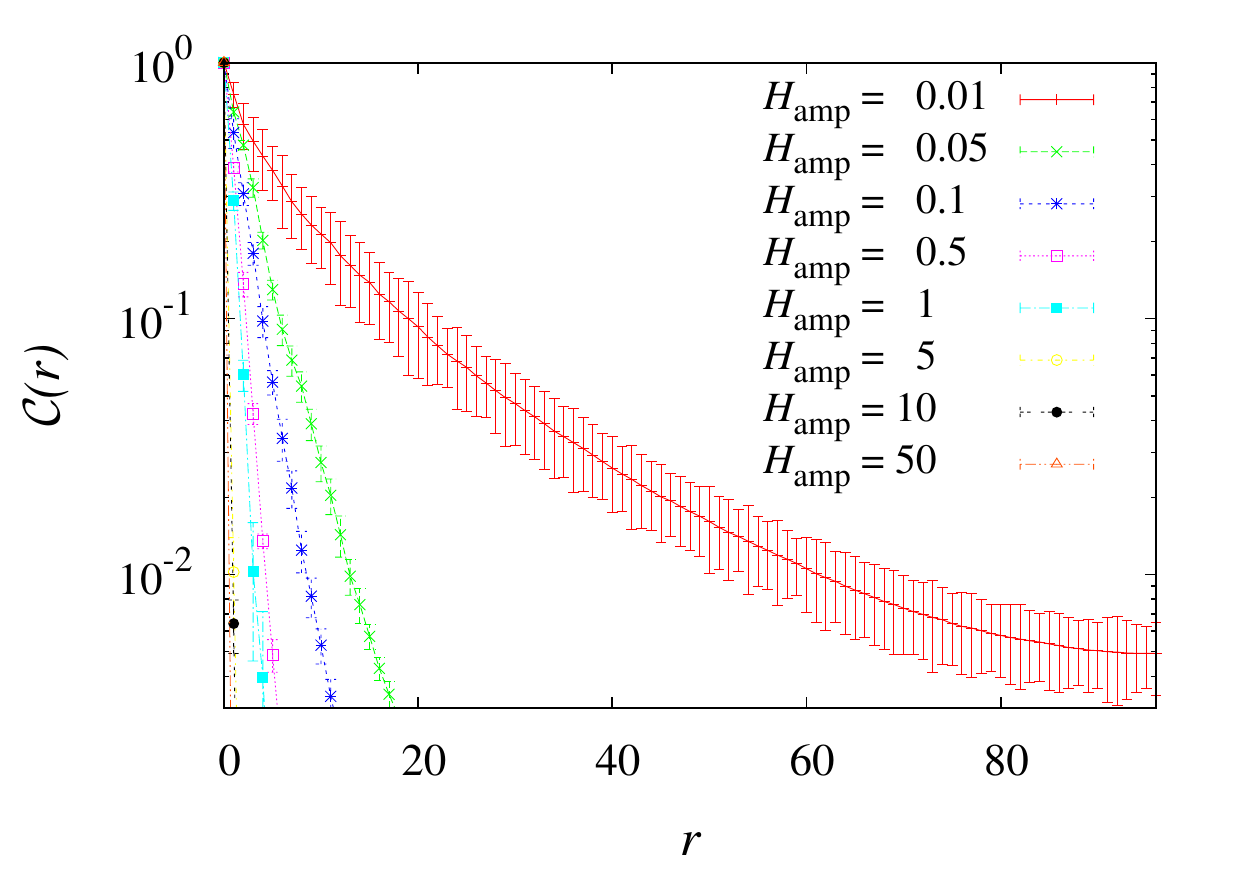}
 \caption{The correlation function $\mathcal{C}(r)$ extracted from
 the eigenvectors in $L=192$ lattices.}
 \label{fig:inverso_L192_zoom}
\end{figure}

As we show in Fig. \ref{fig:inverso_L192_zoom}, when $H_\mathrm{amp}$ is small the correlation function
is not exponential.

\section{Forcings}
\subsection{Probing the linear regime}
\begin{figure}[!htb]
 \centering
 \includegraphics[width=0.48\columnwidth]{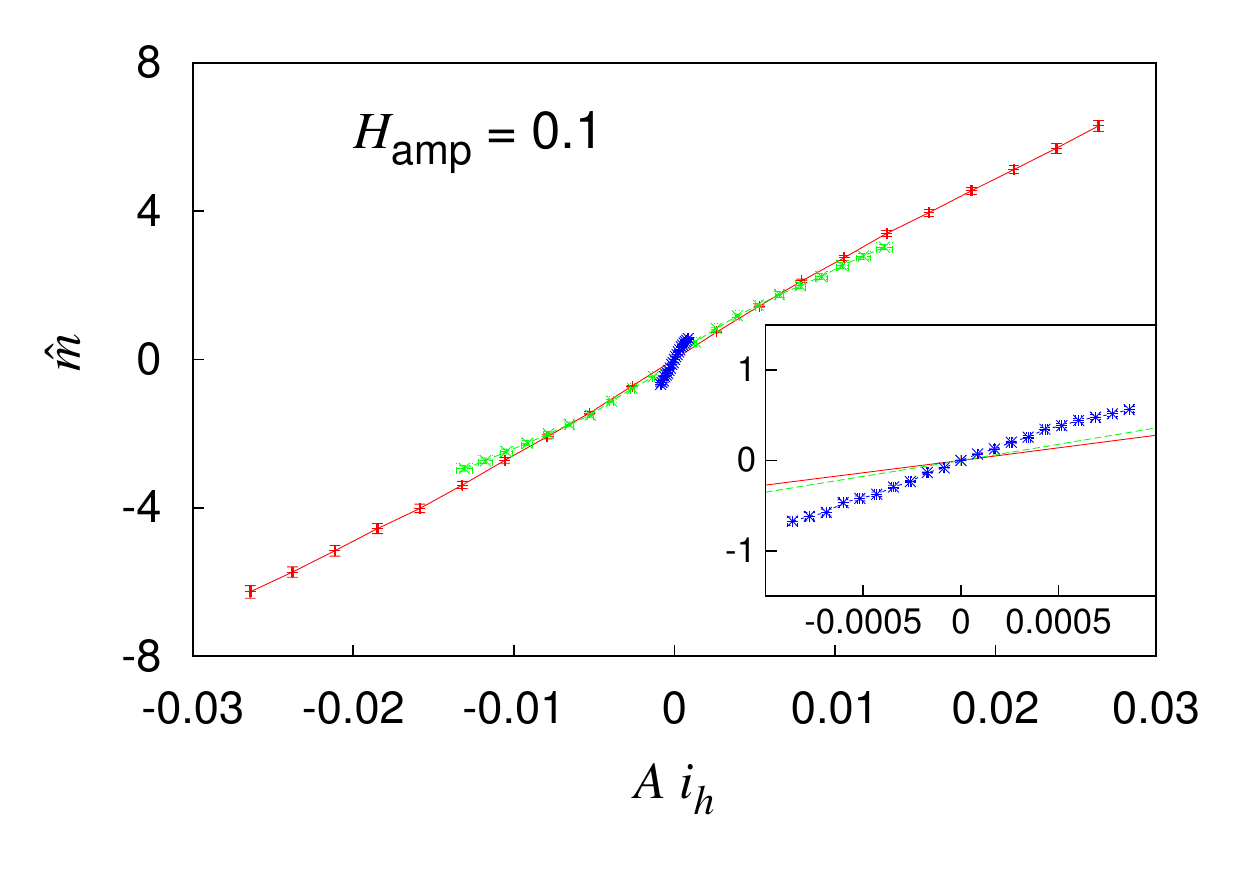}
 \includegraphics[width=0.48\columnwidth]{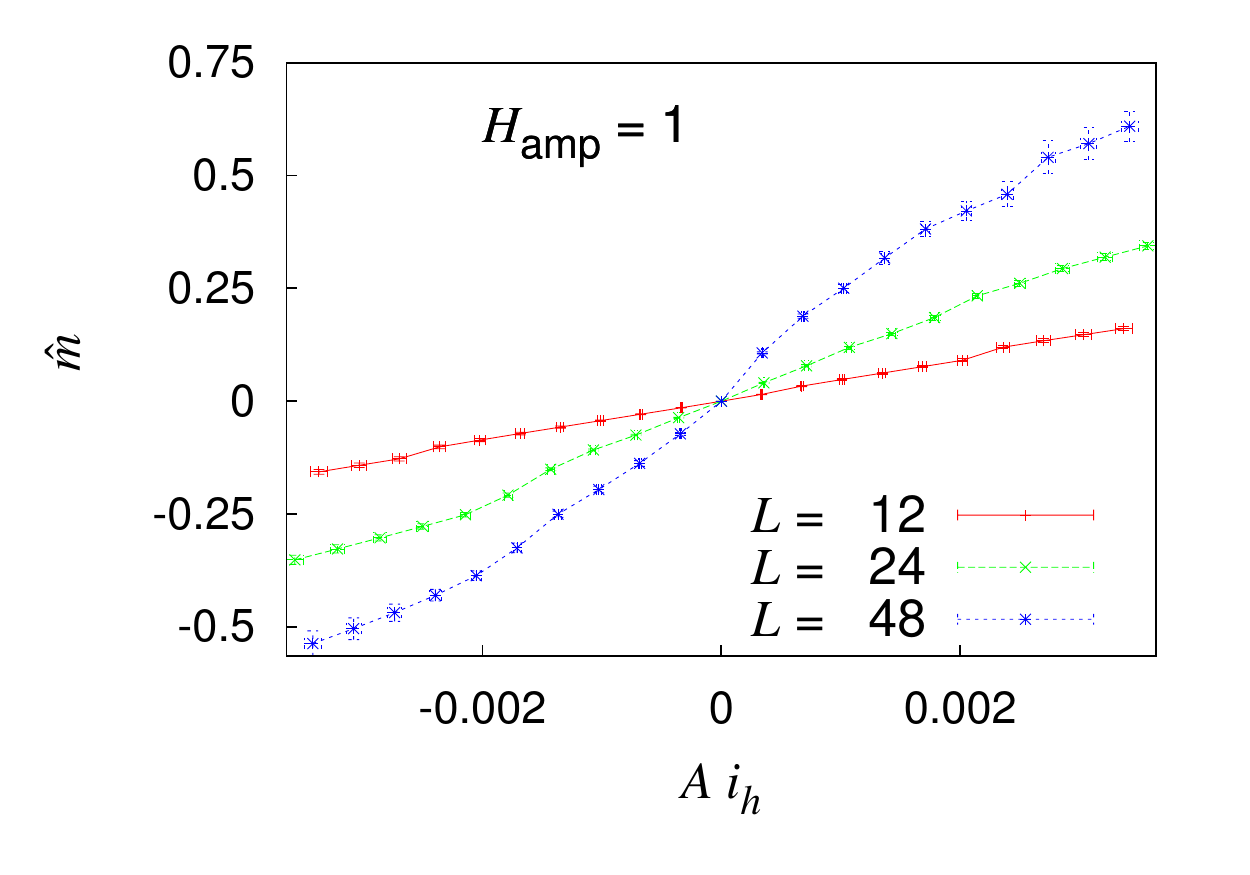}
 \caption[Polarized magnetization $\hat{m}$ of the forcings along $\ket{\vec\pi_0}$]{Polarized magnetization $\hat{m}$ of the forcings along $\ket{\vec\pi_0}$,
 for $H_\amp=0.1$ (\textbf{left}) and $H_\amp=1$ (\textbf{right}). The \textbf{inset} is a zoom of the same data.}
 \label{fig:hatm-pi0}
\end{figure}

\begin{figure}[!htb]
 \centering
 \includegraphics[width=0.48\columnwidth]{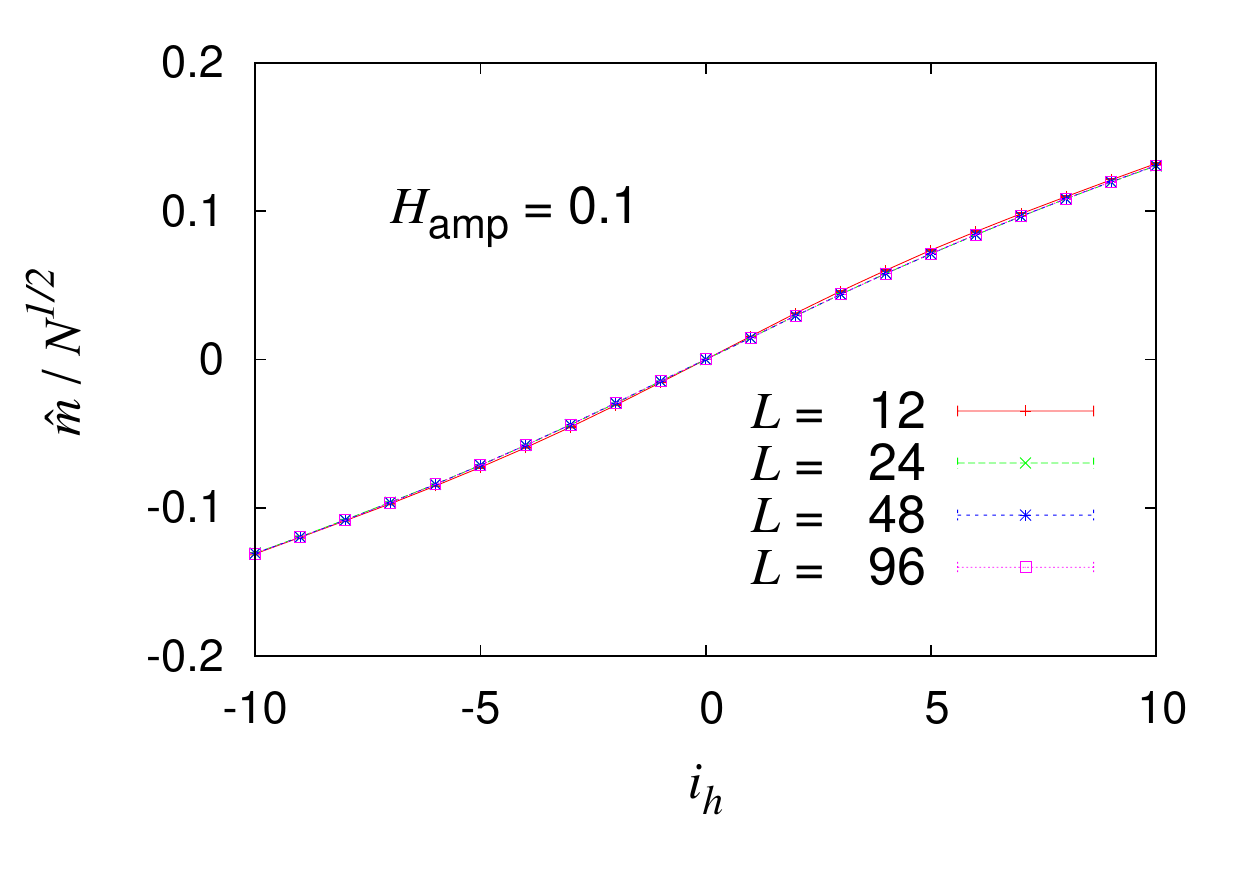}
 \includegraphics[width=0.48\columnwidth]{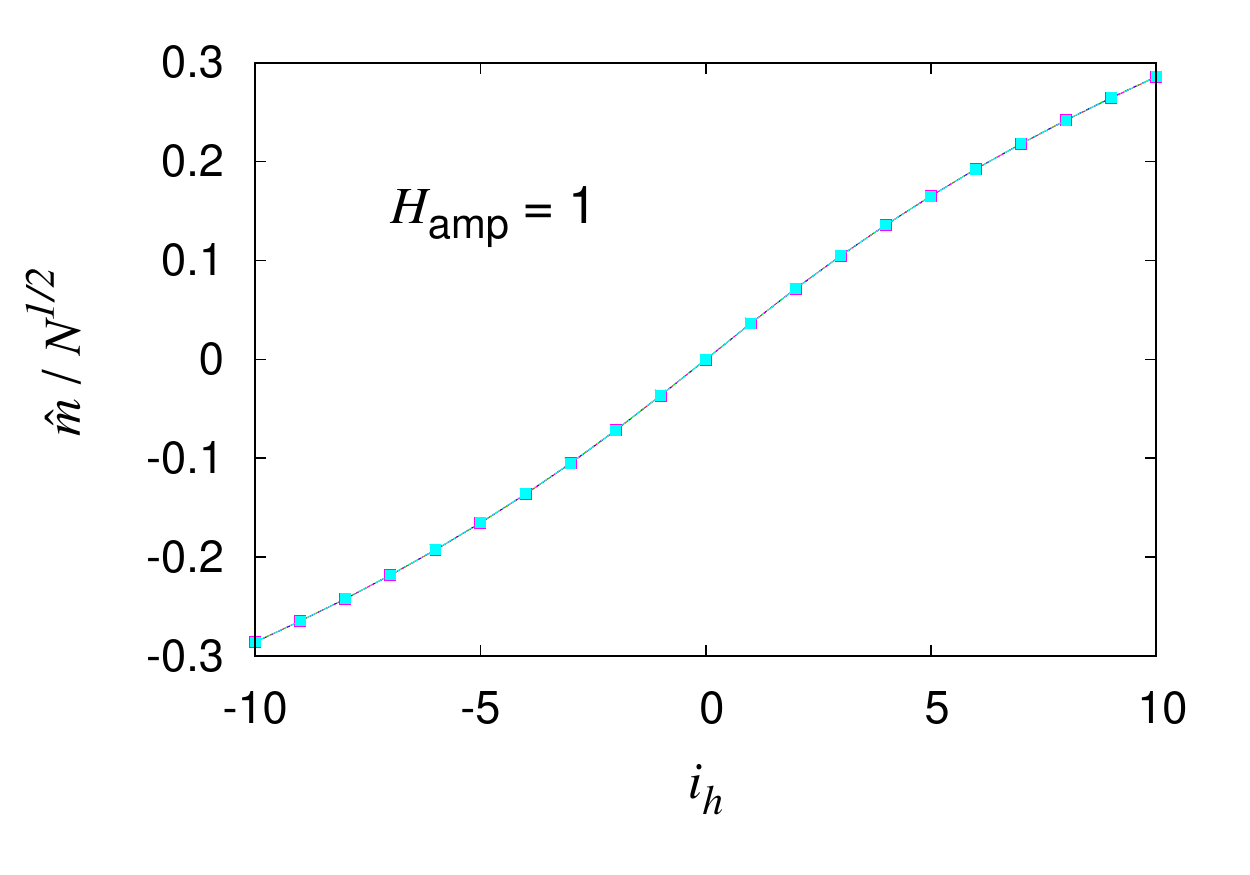}
 \caption[Recaled polarized magnetization $\hat{m}$ of the forcings along $\ket{\vec\pi_\RAND}$]{Rescaled polarized magnetization $\hat{m}$ of the forcings along $\ket{\vec\pi_\RAND}$,
 for $H_\amp=0.1$ (\textbf{left}) and $H_\amp=1$ (\textbf{right}). The data are rescaled in order to collapse.}
 \label{fig:hatm-piRAND}
\end{figure}

To make sure that our forcings are not too strong, we monitor the direct reaction of the system
to the forcing.
We define a 
``polarized magnetization'' \,$\hat{m} = \langle\IS(i_h)\ket{\vec\pi} = \sum_\bx \vec s_\bx\cdot\pi_{\bx}$, that indicates
how much the forcing pushed the alignment of the spins along the pion.
The amplitude of the forcing is tuned well if $\hat{m}(i_h)$ is close to the linear regime.
In table \ref{tab:hsgrf-sim} we show the amplitudes $A$ we used in order to be in the linear regime.
Figure \ref{fig:hatm-pi0} confirms that this was the working condition for the forcings along $\ket{\vec\pi_0}$. Figure \ref{fig:hatm-piRAND} is analogous,
but along $\ket{\vec\pi_\RAND}$. In the latter figure we rescale $\hat{m}$ by a factor $1/\sqrt{N}$ to obtain a collapse. In fact the normalization $\langle \vec\pi_\RAND\ket{\vec\pi_\RAND}=1$
implies that the components of $\ket{\vec\pi_\RAND}$ are of order $1/\sqrt{N}$, so the polarized magnetization is bounded by
$|\hat{m}| = |\langle IS(i_h)\ket{\vec\pi_\RAND}| \leq \sum_\bx |\vec\pi_\bx| \sim \sqrt{N}$.

\paragraph{Ending in a new valley.}
For each $A_\F(i_h)$ we measure the overlap $q_\mathrm{if}$ 
between the two minimas of $\mathcal{H}_\RF$, the initial IS,
$\ket{\vec{s}^{(\,\IS)}}$, and the final one, $\ket{\IS^*}$. Na\"ively, checking that $q_\mathrm{if}<1$ in principle is a good
criterion to establish whether the system escaped to another valley. We proceeded similarly, in terms of the spin
variations between initial and final configuration, through the quantities
\begin{align}
\label{eq:wx}
 w_\bx       =& 1 - \left(\vec{s}^{\,(\IS)}_\bx \cdot \vec{s}^{\,(\IS)\,*}_\bx\right)\,,\\
\label{eq:W}
 W           =& \sum_\bx^N w_\bx = N - \langle\vec{s}^{\,(\IS)}\ket{\IS^*} = N (1-q_\mathrm{if}) \,,\\
\label{eq:w}
 \mathcal{W} =& \frac{\sum_\bx^N w_\bx^2}{\left(\sum_\bx^N w_\bx\right)^2}\,.
\end{align}
The local variation $w_x$ measures the change between the beginning and the end of the process. If the spin
stayed the same then $w_x=0$, while if it became uncorrelated with the initial position $w_x=1$ in average.
If one and only one spin becomes uncorrelated with its initial configuration, the variation of $W$ is $\Delta W=1/N$.
Similar variations $\Delta W$ do not mean that one spin has decorrelated and the others have stayed the same, this
is impossible because $\ket{\vec{s}^{(\IS)}}$ and $\ket{\IS^*}$ are ISs and collective rearrangements are needed. 
A $\Delta W=1/N$ means instead that the overall change is equivalent to a single spin becoming independent of its 
initial state.
This is, for a rearrangement, the minimal change in the $W$ that we can define. Since the spins in our model
are continuous variables, we impose $\Delta W = 1/N$ as a threshold to state whether there was or not
a change of valley.

The cumulant $\mathcal{W}$ is an indicator of the type of rearrangement that took place. If the rearrangement is completely localized
(only one spin changes), $\mathcal{W}=1$, whereas if it is maximally delocalized (all the spins have the same variation), then $\mathcal{W}=1/N$. 

\subparagraph{Falling back in the same valley.}
Even though the forcing is along a definite direction, since the energy landscape is very irregular,
it may happen that stronger forcings lead the system to the original valley. For example it may happen that
$i_h=2$ lead the system to a new valley, and $i_h=3$ lead it once again to the same valley of $i_h=1$. To exclude
these extra apparent valleys we label each visited valley with its $W$, and assume that two valleys with the same 
label are the same valley.
These events are not probable, and even less likely it is that this happen with two different but
equally-labelled valleys, so we neglect the bias due to this unlucky possibility.

\subsection{Rearrangements}
To delineate the shape of the energy landscape, we want to study, for every couple $(H_\amp,L)$, 
the probability that a forcing of amplitude $A_\F$ lead the system to a new valley, 
to distinguish the behavior of soft from bulk modes.

Furthermore, once the system made its first jump to a new valley, it is not excluded that a bigger forcing
lead it to a third minimum of the energy. One can ask himself what is the probability $P_{H_\amp,L}(A_\F, n)$
that $n$ new valleys are reached by forcing the system with an amplitude up to $A_\F(i_h)$, and to try to evince 
a dependency on sistem size and random field. Even though $n$ is bounded by $i_h$, this does not necessarily mean that if we
made smaller and more numerous forcings $n$ could not be larger.
On another side, if for a certain parameter choice rearrangements are measured only for large $i_h$, it is reasonable to think
that these represent the smallest possible forcings to fall off the IS.

To construct $P_{H_\amp,L}(A_\F, n)$, for every replica and sample
we start from $i_h=0$ and increase $|i_h|$ either in the positive or negative direction (the two are accounted for
independently). The value we assign to $P_{H_\amp,L}(A_\F, n)$ is the number of systems that had $n$
rearrangements after $i_h$ steps, divided by the total number of forcings, that is $2N_\mathrm{rep}N_\mathrm{sam}$.

\begin{figure}[!tbh]
\centering
\includegraphics[width=0.48\columnwidth]{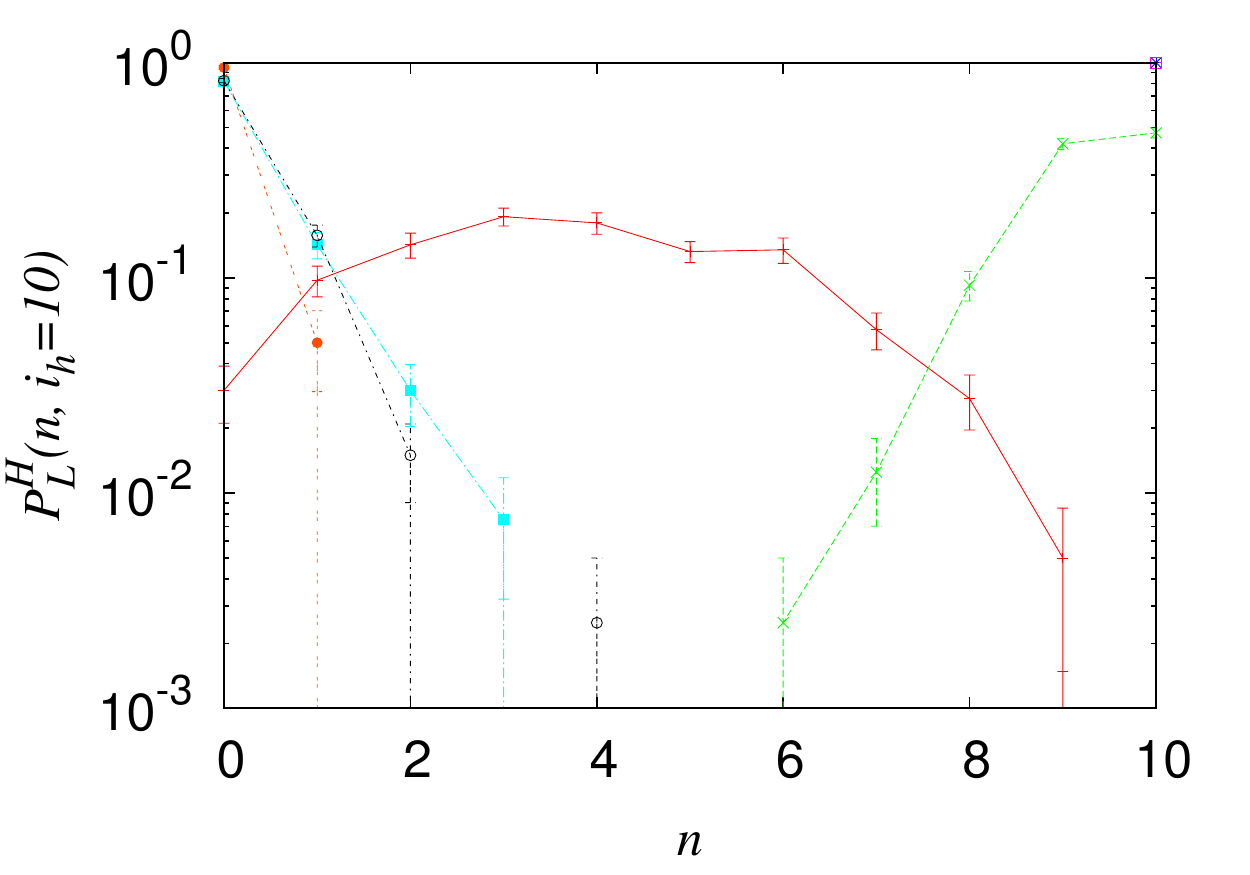}
\includegraphics[width=0.48\columnwidth]{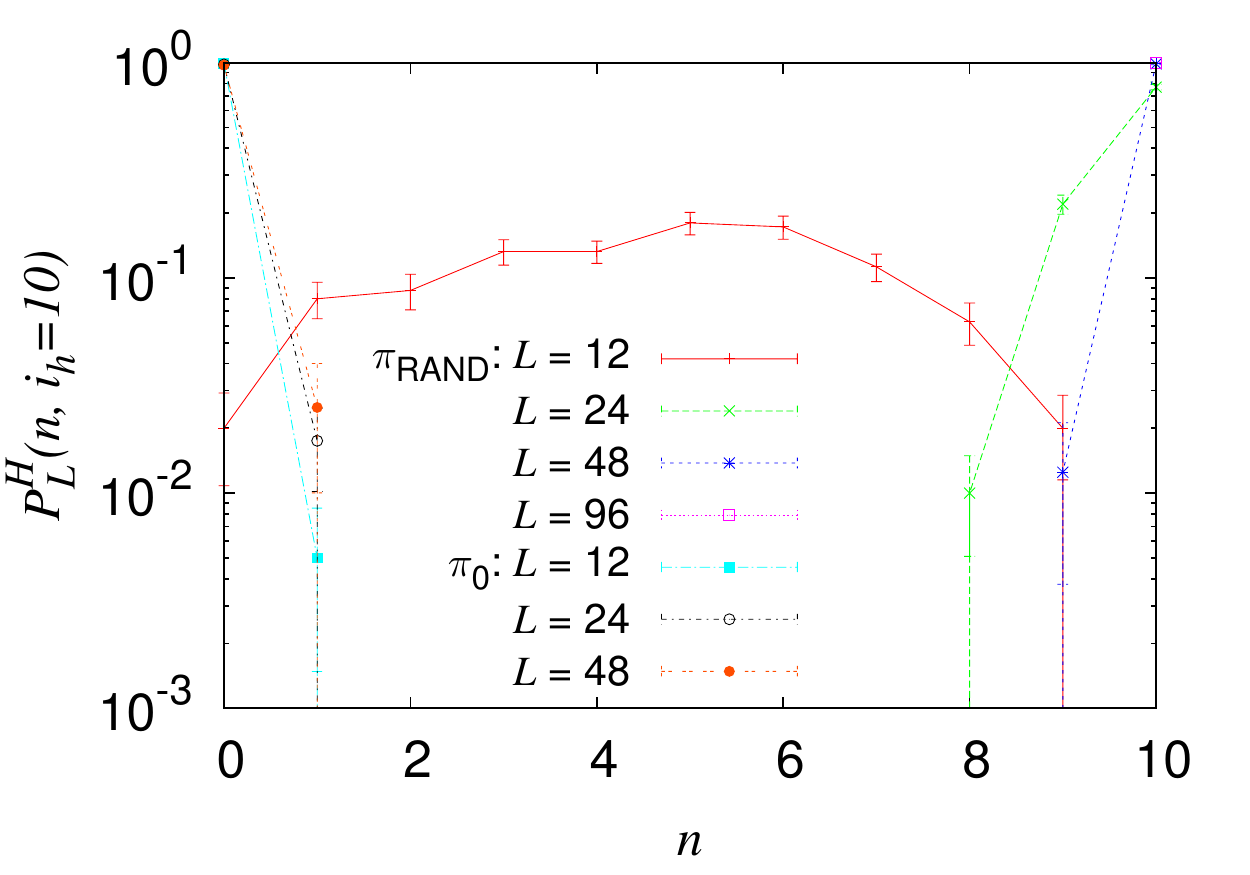}
\caption[Probability of $n$ rearrangements]{Probability of there being exactly $n$ changes of valley after $i_h=N_\F=10$ forcing steps.
 The data come from $H_\amp=0.1$ (\textbf{left}) and $H_\amp=1$ (\textbf{right}).
 If $P_{H_\amp,L}(A_\F, n)=1$ for $n=0$ it means that the forcings were not strong enough to ever get out of the initial IS. On the 
 contrary, $P_{H_\amp,L}(A_\F, n)=1$ for $n=10$ means that every single step lead the system to a new IS. The latter scenario is realized
 in the case of forcings along $\ket{\vec\pi_\RAND}$, especially when the system size is large.
 On the other side, forcings along $\ket{\vec\pi_0}$ display a small but finite amount of rearrangements.}
\label{fig:first-rearrangement-1}
\end{figure}

\begin{figure}[!htb]
\centering
\includegraphics[width=0.48\columnwidth]{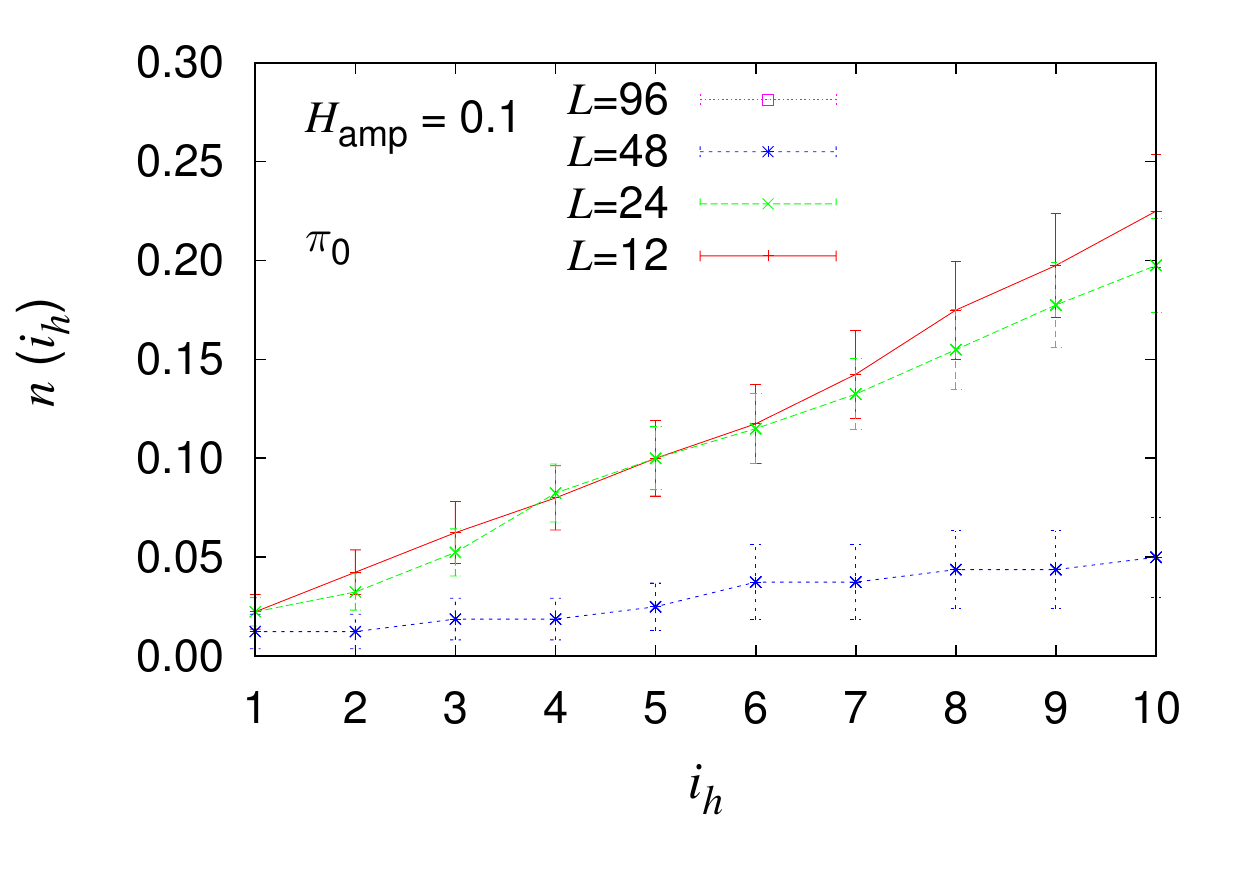}
\includegraphics[width=0.48\columnwidth]{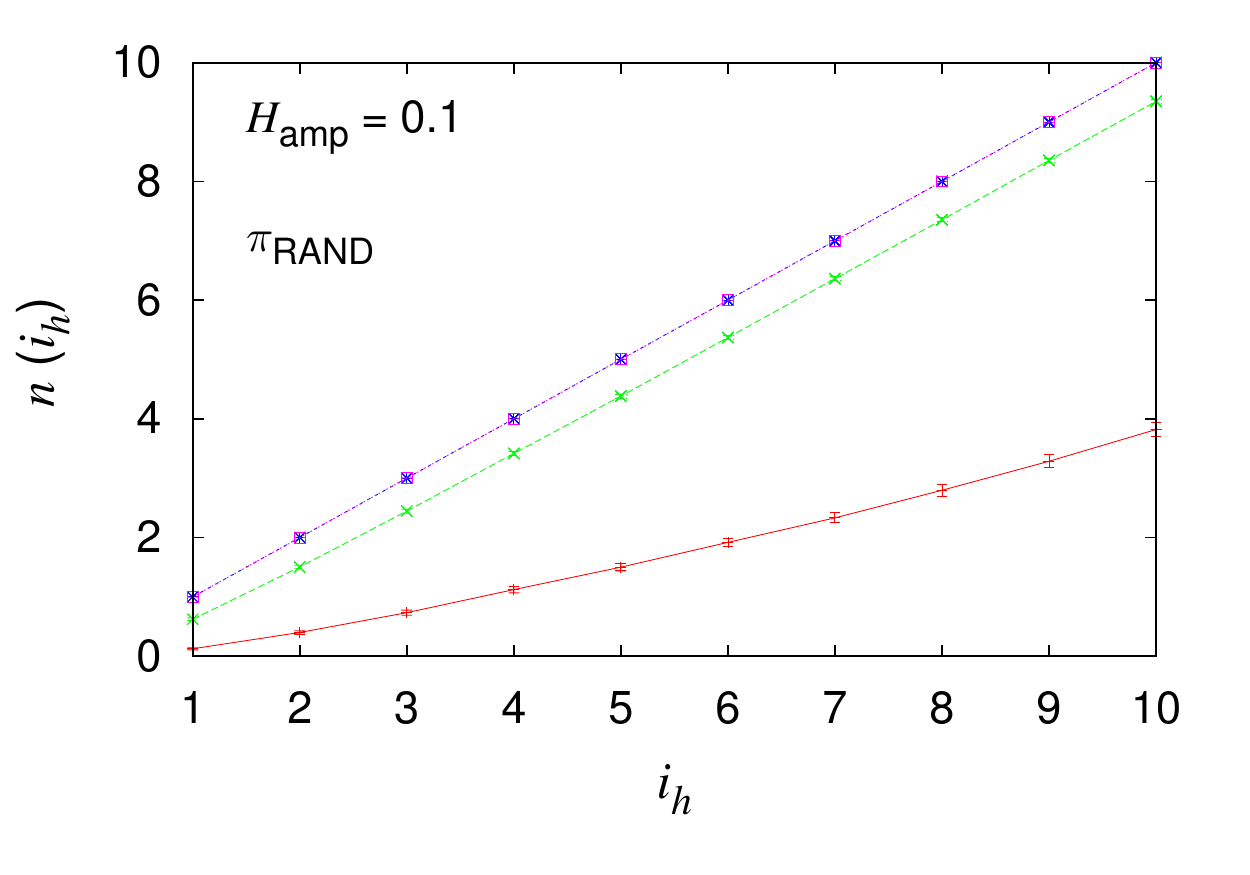}
\includegraphics[width=0.48\columnwidth]{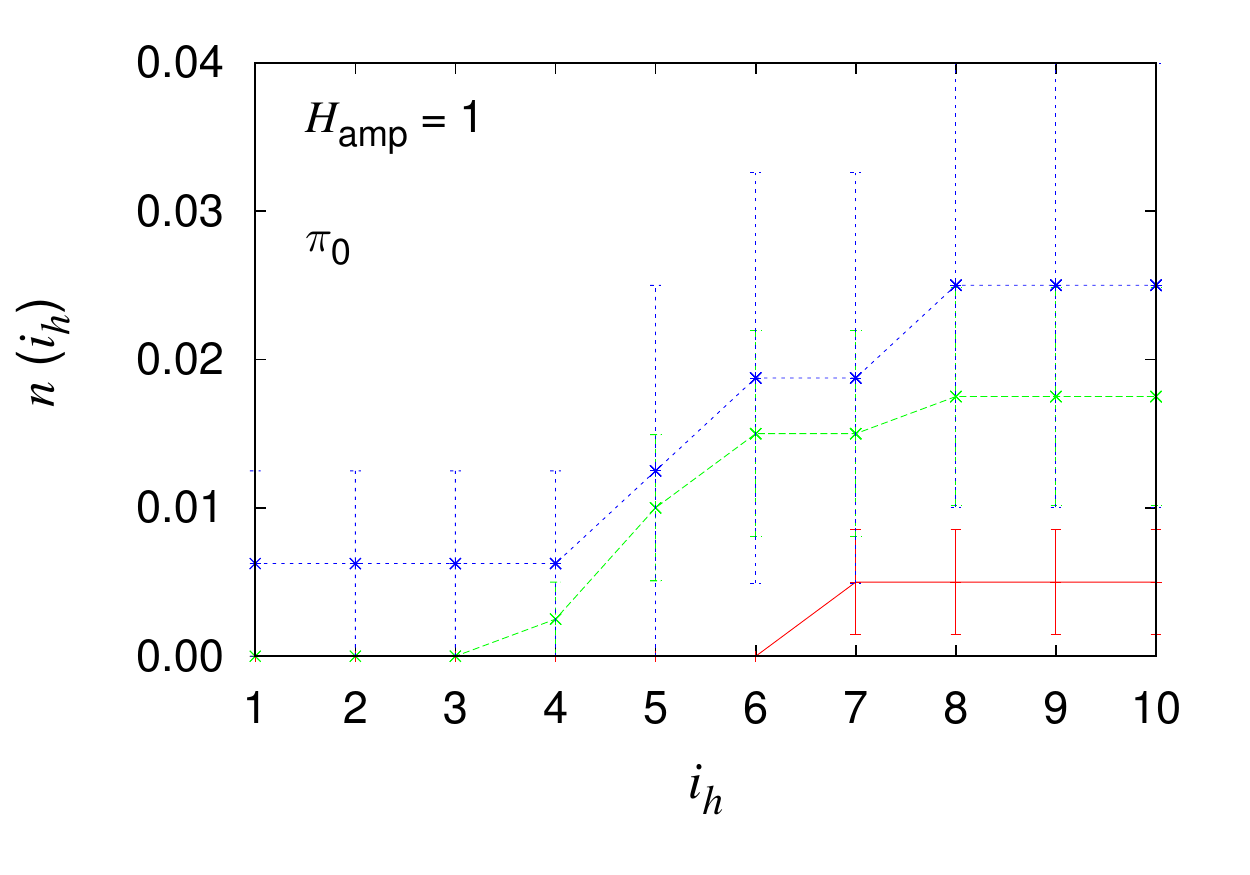}
\includegraphics[width=0.48\columnwidth]{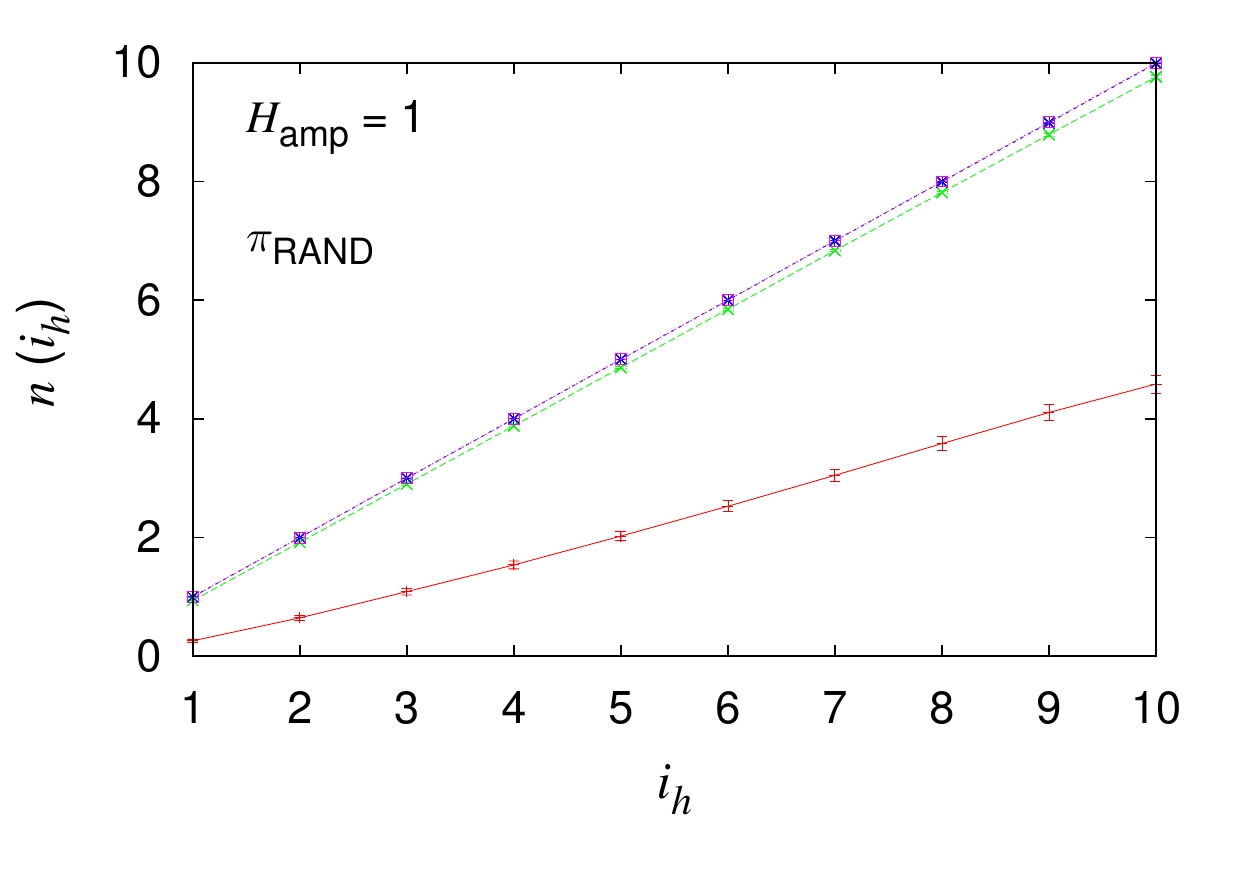}
\caption[Average number of rearrangements]
	{Average number of rearrangements $n(i_h)$ for forcings along $\ket{\vec\pi_0}$ (\textbf{left}) and along $\ket{\vec\pi_\RAND}$ (\textbf{right}). 
	The data come from $H_\amp=0.1$ (\textbf{top}) and $H_\amp=1$ (\textbf{bottom}). 
	When the lattice becomes large enough, the forcings along $\ket{\vec\pi_\RAND}$ lead to a new IS every time $i_h$ is increased.
	The data from the $\ket{\vec\pi_0}$ and $H_\amp=1$ can be said to be in the regime
	of first rearrangement.
	}
\label{fig:first-rearrangement-2}
\end{figure}

\paragraph{First rearrangement.}
In figure \ref{fig:first-rearrangement-1} we show the probability of measuring exactly $n$ rearrangements after
$i_h=N_\F=10$ forcing steps. 
\footnote{We do not show data regarding forcings for $H_\amp=10, 50$, because no arrangement takes place. 
Most likely the energy landscape is too trivial.}
Even though both for $\ket{\vec\pi_\RAND}$ and $\ket{\vec\pi_0}$ we are in the linear response regime, the behavior
is very different between the two types of forcing. In the first case every single forcing step we impose leads the system
to a new valley. In the second rearrangements are so uncommon that even though the probability of having exactly one rearrangement is
finite, that of having more than one becomes negligible for large samples.
It is then reasonable to think that any rearrangement we measure for $\ket{\vec\pi_0}$, it occurs for the smallest possible forcing, and
even when more than one occurs, these jumps are between \emph{neighboring valleys}, where by neighboring we mean that no smaller forcing
would lead the system to a different IS.
To convince ourselves of this we can give a look at the average number of rearrangements after $i_h$ forcing steps, $n(i_h)$ (figure \ref{fig:first-rearrangement-2}). 
\footnote{
Because $P_{H_\amp,L}(A_\F, n)$ is not defined over all the samples
(it is hard to reach many different valleys and it may not happen in all the simulations),
the errors on $P_{H_\amp,L}(A_\F, n)$ were calculated by resampling over the reduced data sets
with the bootstrap method.
}
When $i_h$ is small no new ISs are visited and $n(i_h)=0$, while for larger $i_h$, $n(i_h)$ is positive but small, so we can call
these changes of valley ``first rearrangements'', i.e. rearrangement between \emph{neighboring valleys}.

\subsection{Energy Barriers}
\begin{figure}[!htb]
  \includegraphics[width=\columnwidth]{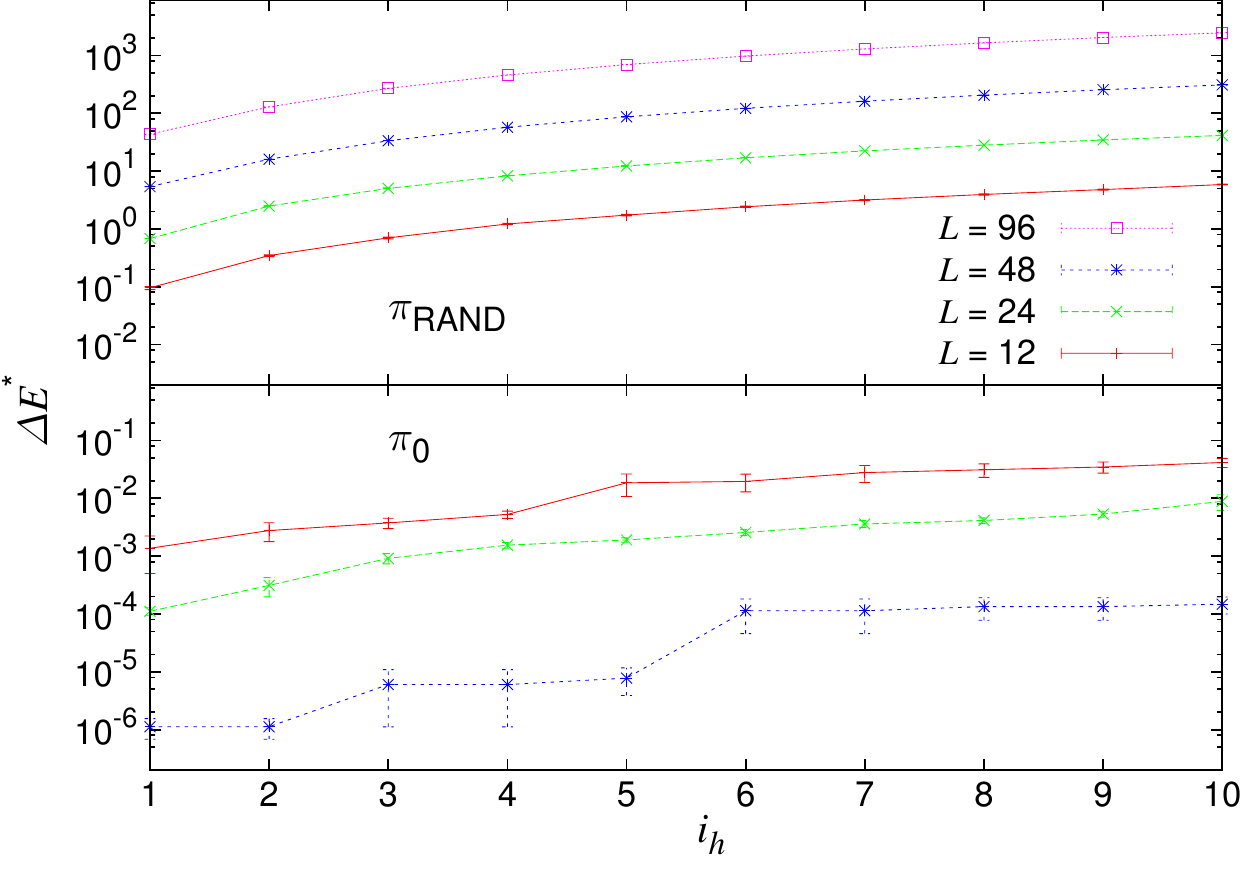}
 \caption{Average energy barrier $\Delta E^*$ for forcings along $\ket{\vec\pi_\RAND}$ (top) and $\ket{\vec\pi_0}$ (bottom), for random fields of amplitude 
	  $H_\amp=0.1$ (data for $\ket{\vec\pi_0}$ and $L=96$ were not computed, see table I).}
\label{fig:energy}
\end{figure}

To reinforce the idea of two-level system (TLS), we report the energy barriers between the couples of connected ISs.
The maximum value of $\Delta E$ before a hop to another valley should give an estimate of height of the barrier. 
Still, it may happen that the IS obtained with the Hamiltonian $\mathcal{H}_\F$ have an energy 
lower than the energy $E_\RF\big(\ket{\vec{s}^{(\,\IS)}}\big)$ calculated with $\mathcal{H}_\RF$, so in a strict sense $\Delta E$ is not positive definite.
To overcome this issue, we measure the barrier $\Delta E^*$ from the arriving IS instead of the starting one, that has 
the advantage of being positive definite.
We see in figure \ref{fig:energy} that while the energy barriers from random forcings increase
with the system size (the growth is $O(N)$), while those within the TLS (along the softest mode) do not grow at all.

\bibliographystyle{unsrt}
\bibliography{biblio.bib}

\end{document}